%% file: main.tex
\documentclass[sigconf,screen,nonacm]{acmart}

\usepackage{tikz}
\usepackage{amsmath}

\usepackage{filecontents}

\usepackage{amsmath}
\usepackage{amsthm}
\usepackage{amssymb}
\usepackage{mathtools}
\usepackage{bm}
\usepackage{bbm}
\usepackage{booktabs}
\usepackage{xspace}
\usepackage{color,soul}
\usepackage{multirow}
\usepackage{multicol}
\usepackage{caption}
\usepackage{enumitem}
\usepackage{verbatim}
\usepackage{microtype}
\usepackage{graphicx}
\usepackage{booktabs}
\usepackage{hyperref}
\usepackage{nicefrac}
\usepackage[normalem]{ulem}
\usepackage{tikz}
\usetikzlibrary{trees}
\usepackage{float}
\usepackage{subcaption}
\usepackage[ruled,noend,linesnumbered,vlined]{algorithm2e}
\usepackage{xcolor,colortbl}
\usepackage{hhline}
\usepackage{tabularx}
\usepackage{pifont}
\usepackage{circledtext}
\usepackage{makecell}
\usepackage{multirow}
\usepackage{soul}
\usepackage{xcolor}

\usepackage{tikz}
\usepackage{mdframed}
\usetikzlibrary{shadows}
\newmdenv[
  shadow=false,
  font=\small,
  backgroundcolor=white,
  leftmargin=0pt,
  rightmargin=0pt,
  skipabove=2pt,
  skipbelow=2pt,
  innerleftmargin=5pt,     
  innerrightmargin=5pt,
  splittopskip=\topskip,   
  splitbottomskip=\topskip 
]{shadedbox}

\usepackage{listings}
\newcommand{\specialcell}[2][c]{%
	\begin{tabular}[#1]{@{}c@{}}#2\end{tabular}}

\definecolor{lightgreen}{rgb}{0.8, 1.0, 0.8}

\newcommand{\reviewerB}[1]{{#1}}

\newcommand{\reviewerA}[1]{{#1}}
\newcommand{\reviewerC}[1]{{#1}}

\newcommand{\myvspace}[1]{\vspace{#1}}

\definecolor{myred}{rgb}{1.0,0.7,0.8}
\definecolor{mygreen}{RGB}{0,166,0}
\definecolor{lightgreen}{rgb}{0.56, 0.93, 0.56}
\definecolor{myorange}{RGB}{252,107,4}
\definecolor{darkgreen}{RGB}{0,153,102}
\definecolor{lightblue}{rgb}{0.53, 0.81, 0.92}
\definecolor{lightgray}{gray}{0.9}

\newcommand{\mysubsubsection}[1]{{\vspace{0.25em}\noindent\ul{\textbf{\textit{#1.\xspace}}}}}
\newcommand{\system}{{\textsc{StaleFlow}}\xspace}

\newtheorem*{assumption*}{\assumptionnumber}
\providecommand{\assumptionnumber}{}


\newtheorem*{statement*}{\statementnumber}
\providecommand{\statementnumber}{}


\newcommand\footnotewithoutmarker[1]{%
  \begingroup
  \renewcommand\thefootnote{}\footnote{#1}%
  \addtocounter{footnote}{-1}%
  \endgroup
}

\begin{document}

\title{\system: Staleness-Aware Data Management for Mitigating Data Skewness in Fully Disaggregated RL Post-Training}

\author{Haoyang Li{$^{*1}$}, Sheng Lin{$^{*1}$}, Fangcheng Fu{$^{2}$}, Yuming Zhou{$^{1}$}, Xiaodong Ji{$^{1}$}, Yanfeng Zhao{$^{2}$}, Lefeng Wang{$^{2}$}, Jie Jiang{$^{3}$}, Bin Cui{$^{1}$}}

\affiliation{
{$^{1}$}Peking University\country{China} \hspace{1.2em}{$^{2}$}Shanghai Jiao Tong
University\country{China} \hspace{1.2em}
{$^{3}$}Tencent\country{China}\\
}

\begin{abstract}
Reinforcement learning (RL) post-training has become pivotal for enhancing the capabilities of modern large models. A recent trend is to develop RL systems with a fully disaggregated architecture, which decouples the three RL phases (rollout, reward, and training) onto separate resources and executes them asynchronously. However, two critical data-level concerns arise: (1) asynchronous execution leads to \textit{data staleness} in trajectories (the data generated by rollout) as the model parameters used in rollout may not be up to date, which impairs RL convergence; and (2) the length variation of trajectories introduces severe \textit{data skewness}, leading to workload imbalance and degraded system performance.

Existing systems fail to address these two concerns in a unified manner. Techniques that tightly control data staleness often constrain effective data skewness mitigation, while aggressive data skewness mitigation tends to exacerbate data staleness. As a result, systems are forced to trade off convergence for performance, or vice versa. To address this, we propose \system, an RL post-training system that jointly tackles data staleness and skewness. First, to control staleness, \system introduces a global consistency protocol that tracks the full lifecycle of each trajectory and constrains staleness. Second, to mitigate skewness, \system re-designs the RL system architecture by constructing data servers for trajectories and parameters to achieve flexible rollout coordination. Subsequently, we develop a suite of staleness-aware, throughput-oriented strategies to enhance system performance. 
Evaluations show that \system achieves up to 1.42-2.68$\times$ (\reviewerC{1.18-1.91}$\times$ on average) higher throughput than state-of-the-art systems, without compromising convergence. Our source code is available: {\small \url{https://github.com/psrl-project/psrl}}.
\footnotewithoutmarker{$^{*}$Equal contribution}
\footnotewithoutmarker{Contact: Haoyang Li (2000012918@stu.pku.edu.cn), Fangcheng Fu (ccchengff@sjtu.edu.cn) and Bin Cui (bin.cui@pku.edu.cn)}
\end{abstract}

\maketitle
\pagestyle{plain}

\input{sections/intro}

\input{sections/background}

\input{sections/overview}

\input{sections/staleness}

\input{sections/skewness}

\input{sections/exp}

\input{sections/conclusion}

\begin{acks}
This work is supported by National Natural Science Foundation of China (U23B2048, 62402011), Fundamental and Interdisciplinary Disciplines Breakthrough Plan of the Ministry of Education of China (JYB2025XDXM108), Beijing Municipal Natural Science Foundation (Grant No. L2603008), the Fundamental Research Funds for the Central Universities, Peking University, and PKU-Tencent joint research Lab. Fangcheng Fu and Bin Cui are the corresponding authors.
\end{acks}

\bibliographystyle{plain}
\bibliography{reference}

\clearpage
\appendix
\input{sections/appendix}


\end{document}

%% file: sections/intro.tex
\myvspace{-1pt}
\section{Introduction}
\label{sec:intro}

With the diminishing returns from scaling large model pre-training, the research community has increasingly focused on scaling reinforcement learning (RL) during post-training as the next frontier for advancing model capabilities~\cite{scale_rl_1, scale_rl_2}. Recently, we have witnessed the emergence of large models with deep reasoning capabilities (e.g.,  OpenAI-o1~\cite{openai_o1}, Kimi-K2~\cite{kimi_k2}, DeepSeek-R1~\cite{deepseek_r1}), many of which rely on RL to enable long-chain reasoning~\cite{rl_reasoning_1, rl_reasoning_2, rl_reasoning_3}.

A common workflow of RL post-training consists of three phases: \emph{rollout}, \emph{reward}, and \emph{training}~\cite{trpo, ppo, grpo, dapo, gspo}. 
A growing trend is to decouple these phases, deploy them on dedicated resources, and execute them asynchronously. Compared to shared-resource designs, such disaggregated architectures typically offer substantially better scalability~\cite{stream_rl, verl_pipeline, llama_rl, areal, roll_flash, rhy_rl, laminar, async_rlhf, asyncflow}. 

\begin{figure}[!t]
    \centering
    \includegraphics[width=\linewidth]{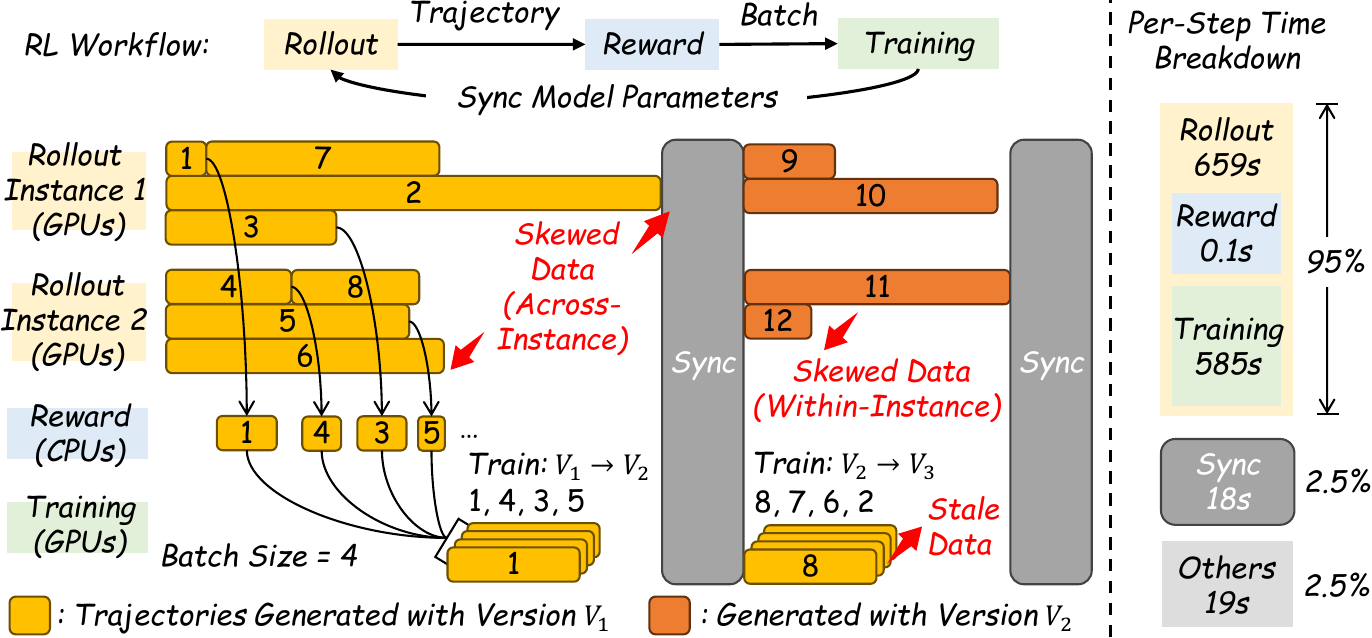}
    \myvspace{-18pt}
    \caption{\small{(Left) RL workflow in a fully disaggregated architecture. (Right) Time breakdown of post-training a Qwen3-30B-A3B (64 H20 GPUs). Rollout and training overlap and jointly dominate the time.}}
    \label{fig:rl_workload}
    \myvspace{-5pt}
\end{figure}

From the data management perspective, there are two kinds of data that move across different phases, namely \textit{trajectories} and \textit{parameters}.
As illustrated in Figure~\ref{fig:rl_workload}, 
RL trajectories are generated in a streaming manner across multiple rollout instances (i.e., model replicas), and then forwarded to the reward stage upon completion. 
Once sufficient rewarded trajectories are accumulated to form a batch, the model is trained and updated. 
Finally, the model parameters are synchronized back to the rollout side, completing an RL post-training step. 
In this loop, rollout and training overlap and together dominate the end-to-end execution time.


\mysubsubsection{Concerns from data}
Within this asynchronous workflow, two fundamental data-level concerns arise. (1) \textbf{\textit{Data staleness.}} Asynchronous execution of
rollout and training significantly improves system throughput, but it also causes the training stage to consume stale trajectories, as the parameters used in rollout may not be up to date. 
Undoubtedly, this risks impaired convergence, as analyzed in \S\ref{subsec:data_staleness}.
(2) \textbf{\textit{Data skewness.}} During rollout, trajectory lengths vary substantially, leading to workload skewness both \textit{within} an instance and \textit{across} instances. This imbalance degrades resource utilization and overall system performance, as analyzed in \S\ref{subsec:data_skewness}.

\textit{To control data staleness,} existing systems adopt different approaches, ranging from strong to weak guarantees: (1) Some systems~\cite{areal, roll_flash} allow users to specify an explicit \textit{staleness bound} $\eta$, thereby strictly bounding staleness. (2) Other systems~\cite{verl_pipeline, asyncflow, rhy_rl} do not support configurable bounds and only permit one-step asynchrony (i.e., $\eta=1$). (3) The remaining systems~\cite{llama_rl, laminar, april, sorted_rl} provide no explicit staleness guarantees. \textit{To mitigate data skewness,} various rollout coordination techniques have been proposed: (1) Partial rollout~\cite{llama_rl, ulorl, areal, kimi_partial_rollout, roll_flash, asyncflow}, which allows a single trajectory to be generated by multiple model versions; (2) Redundant rollout~\cite{april, sorted_rl, roll_packer}, which oversamples trajectories and discards long-tail ones; (3) Rollout with multi-version instances~\cite{laminar, rhy_rl, asyncflow}, where instances update models at their own pace without global synchronization; and (4) Migration across instances~\cite{seer, laminar, roll_packer, rhy_rl}, which dynamically redistributes trajectories to rebalance workload.

\begin{figure}[!t]
    \centering
    \includegraphics[width=\linewidth]{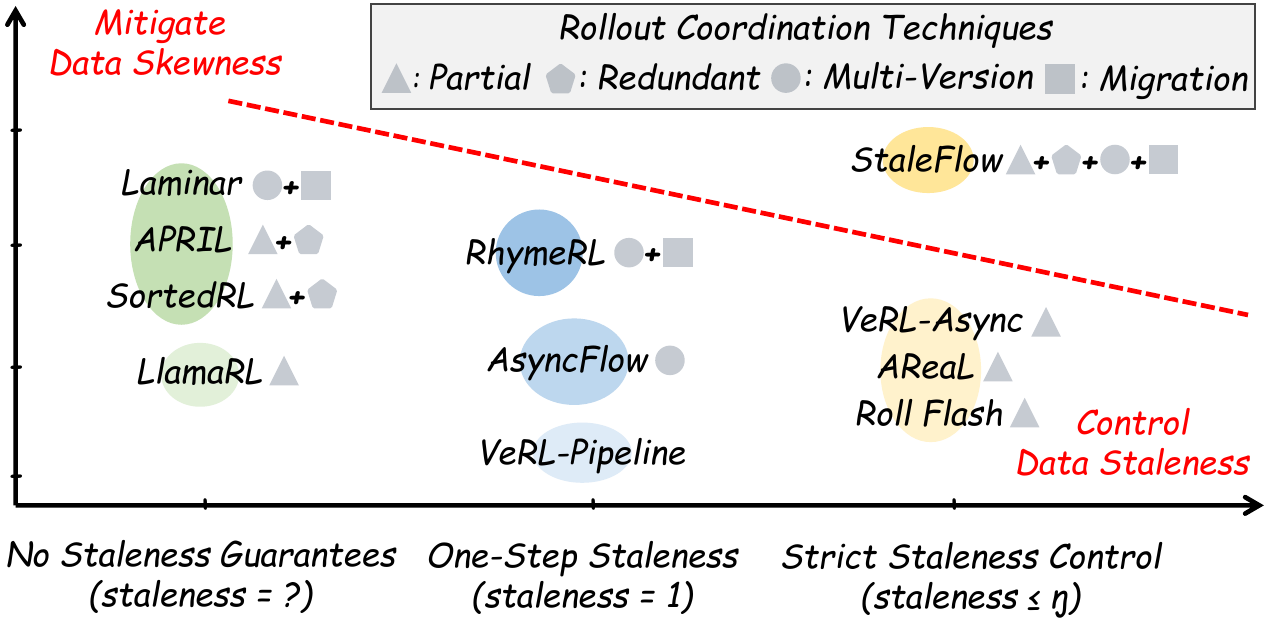}
    \myvspace{-20pt}
    \caption{\small{Comparison of different RL systems.  Higher values on the vertical axis indicate stronger support for rollout coordination.}} 
    \label{fig:compare_sys}
    \myvspace{-5pt}
\end{figure}

\mysubsubsection{Limitations}
Despite these advances, current systems fall short in striking a good balance between staleness control and skewness mitigation, as shown in Figure~\ref{fig:compare_sys}. 
Designs that enforce strict staleness guarantees necessarily constrain skewness mitigation, as many rollout coordination techniques risk violating the \textit{staleness bound}. In contrast, systems that support rich and flexible rollout coordination typically relax staleness control in favor of higher throughput. As a result, existing approaches are forced into a trade-off between RL convergence and system performance.

\mysubsubsection{Challenges and motivations} This limitation arises from two unresolved challenges: 
(1) First, as rollout coordination techniques mostly operate on the trajectories, to synergize both worlds, a fine-grained, trajectory-level staleness control is expected, yet this is unexplored in current strict control approaches. 
(2) Second, existing systems lack a globally coordinated, staleness-aware rollout coordination framework to mitigate skewness: existing techniques are applied in isolation and ad hoc, without explicitly modeling and optimizing system throughput under a given \textit{staleness bound}.

To address these challenges, we present \system, an RL post-training system that jointly provides strong staleness guarantees and high system performance. \system is part of the PSRL project\footnote{\url{https://github.com/psrl-project/psrl}}.
(1) To control data staleness (\S\ref{sec:control_staleness}), we propose a novel global consistency protocol built on a virtual staleness buffer abstraction, which tracks the lifecycle of each trajectory at fine granularity. 
This makes our system compatible with \reviewerA{flexible} rollout coordination techniques while enabling precise staleness control.
(2) To mitigate data skewness and improve system performance (\S\ref{sec:mitigate_skewness}), we introduce innovations in both rollout architecture and algorithms. Architecturally, we introduce two middleware data servers (i.e., a trajectory server and a parameter server) to decouple data movement from rollout. This decoupling allows rollout instances to selectively determine when and how to fetch and coordinate data, substantially improving flexibility. Algorithmically, we develop a suite of staleness-aware, throughput-oriented rollout coordination strategies. These strategies are orchestrated by a centralized coordinator that continuously analyzes system snapshots and issues commands to actively steer system behavior.

In summary, our contributions are as follows:
\begin{itemize}[noitemsep, topsep=0pt, parsep=0pt, partopsep=0pt, leftmargin=*]
\item We propose a global consistency protocol that enforces strict staleness control at the trajectory level while supporting \reviewerA{flexible} rollout coordination techniques (\S\ref{sec:control_staleness}).
\item We introduce architectural innovations that decouple data movement into dedicated servers, serving as middleware to enable flexible rollout coordination (\S\ref{subsec:arch}).
\item We design a suite of staleness-aware, throughput-oriented rollout coordination strategies that unify diverse techniques to maximize system throughput under explicit staleness constraints, implemented via a snapshot–command cycle (\S\ref{subsec:cycle}, \S\ref{subsec:strategy}).
\item Evaluations on a 128-GPU cluster show that \system achieves 1.42-2.68$\times$ (\reviewerC{1.18-1.91}$\times$ on average) throughput improvement over state-of-the-art RL systems, while preserving convergence (\S\ref{sec:exp}).
\end{itemize}

%% file: sections/background.tex
\section{Preliminaries and Motivations}
\label{sec:background}

In this section, we explore the workflow of RL post-training and  investigate two emerging data-level concerns. We then discuss existing solutions addressing these concerns and analyze their limitations, which ultimately motivate the design of our work.

\subsection{RL Workflow and Trend}
\label{subsec:rl_workload}

\mysubsubsection{RL workflow}
As illustrated in Figure~\ref{fig:rl_workload}, a typical RL post-training workflow consists of three distinct phases. (1)  \textit{\textbf{Rollout phase.}}  
The model samples prompts from a dataset and generates responses token-by-token through auto-regressive decoding. The prompt and its corresponding response are then concatenated to form an RL trajectory.\footnote{For clarity, a prompt alone is considered an \textit{initial trajectory} in this paper.}
(2) \textit{\textbf{Reward phase.}} The system applies rule-based or verifiable reward models~\cite{rlvr_1, rlvr_2} to evaluate generated trajectories and assign scalar scores, which serve as reward signals. 
(3) \textit{\textbf{Training phase.}} 
A selected RL algorithm consumes the collected trajectories and their reward, batches the data for model training, and synchronizes the updated parameters back to the rollout side.

\mysubsubsection{Fully disaggregated architecture}
The three RL phases exhibit distinct resource requirements and computational characteristics. Both the rollout and training phases rely on GPUs for acceleration, but they differ significantly in their nature: the rollout phase is memory-bound, performing token-by-token decoding, while the training phase is compute-bound, involving large-batch forward and backward passes, and scales more efficiently with additional GPUs. In contrast, the reward phase typically runs on CPUs, executing rule-based or verifiable evaluations.
Given these differences, a key trend in RL system design is to \textit{fully disaggregate the three phases} and allocate them to specialized resources~\cite{stream_rl, verl_pipeline, llama_rl, areal, roll_flash, rhy_rl, laminar, async_rlhf, asyncflow}. As shown in Figure~\ref{fig:rl_workload}, unlike synchronous RL systems that rely on shared resources and sequential execution (e.g., VeRL~\cite{verl}), this disaggregated architecture enables the phases to execute in an overlapped and asynchronous manner, allowing each phase to independently optimize its resource utilization and performance.

\mysubsubsection{Data-Level concerns}
However, such a disaggregated architecture faces two critical data-level concerns. 
(1) \textit{\textbf{Data staleness.}} To enable overlapped and asynchronous execution of training and rollout, systems need to train on trajectories generated by older versions of the model. For example, as shown in Figure~\ref{fig:rl_workload}, training from version $V_2$ to $V_3$ consumes data generated by $V_1$, thereby allowing the $V_2$ training phase to overlap with the $V_2$ rollout phase. While this improves resource utilization, it introduces stale data, which may impair RL convergence when the staleness becomes excessive.
(2) \textit{\textbf{Data skewness.}} During rollout, trajectory lengths may vary significantly. As shown in Figure~\ref{fig:rl_workload}, because trajectory generation is distributed across multiple rollout instances (i.e., model replicas), this skewness manifests both \textit{within-instance} (uneven loads from long vs. short sequences within the same instance) and \textit{across-instance} (different instances progressing at different speeds). Both forms of skewness degrade overall RL system performance.

\reviewerA{In the following, we present a detailed analysis of data staleness and skewness in RL, exploring their impacts, and classify existing solutions based on how they address them.}

\begin{figure}[!t]
    \centering
    \includegraphics[width=\linewidth]{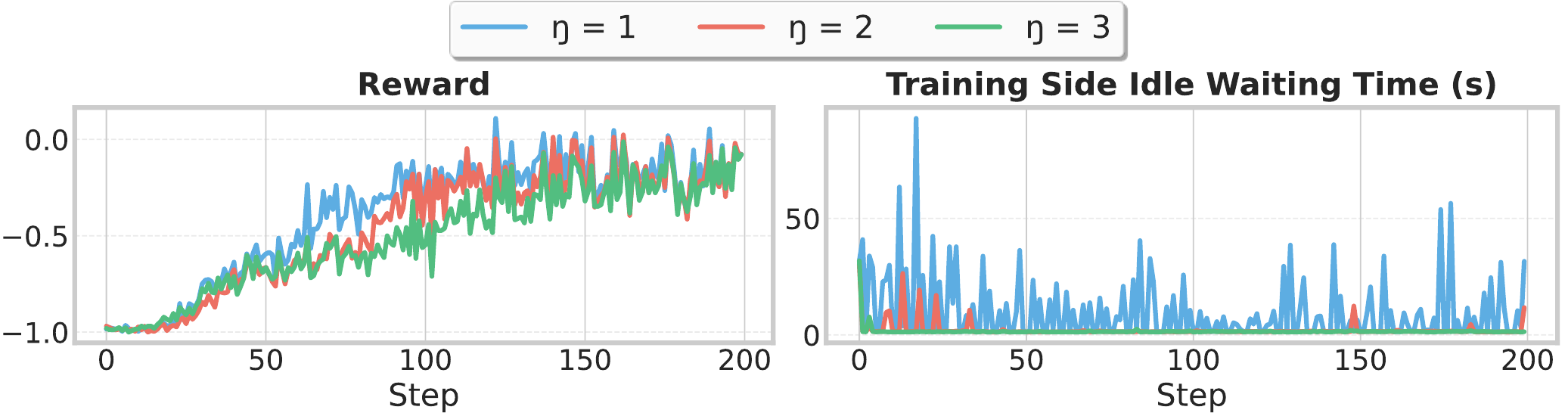}
    \myvspace{-22pt}
    \caption{\small{Effect of \textit{staleness bounds} ($\eta$) on post-training Qwen2.5-MATH-7B with DAPO~\cite{dapo} (64 H20 GPUs). A larger \textit{staleness bound} reduces idle waiting in training but slows convergence.}}
    \label{fig:different_staleness}
    \myvspace{-5pt}
\end{figure}

\subsection{Data Staleness in RL}
\label{subsec:data_staleness}

\mysubsubsection{Impacts}
\reviewerA{To prevent excessive model discrepancy, RL post-training typically imposes a global \textit{staleness bound} $\eta$, defined as the maximum allowed model version lag between the rollout and training phases.} As illustrated in Figure~\ref{fig:different_staleness}, increasing $\eta$ generally degrades convergence, as the rollout model diverges more from the training model. However, tolerating higher staleness can substantially improve system throughput by allowing more rollout trajectories to proceed concurrently and reducing idle waiting times on the training side.


\mysubsubsection{Current solutions}
Existing solutions can be broadly classified into three categories based on how they regulate or tolerate staleness.


(1) \textit{\textbf{Strict staleness control.}} 
The first category, exemplified by VeRL-Async~\cite{verl_async_doc}, AReaL~\cite{areal}, and ROLL Flash~\cite{roll_flash}, grants users full autonomy in selecting $\eta$. They enforce strict staleness control via \reviewerA{simple admission control. Specifically, they assume the model is globally updated (i.e., all rollout instances have the same model version), and maintain a counter of in-flight data (i.e., trajectories being generated or generated but not yet consumed). New trajectories are admitted only if the in-flight data does not exceed the capacity of $(\eta+1)\times\text{batch\_size}$. Otherwise, admission is paused until another batch\_size is consumed by training.} 


(2) \textit{\textbf{One-Step staleness.}} The second category includes systems such as VeRL-Pipeline~\cite{verl_pipeline, verl_pipeline_doc}, AsyncFlow~\cite{asyncflow}, and RhymeRL~\cite{rhy_rl}. These systems do not support user-defined \textit{staleness bounds} and only permit one-step asynchrony (i.e., $\eta=1$). During training, they strictly use a model that is at most one version behind to generate exactly one batch, ensuring that staleness never exceeds one step.


(3) \textit{\textbf{No staleness guarantees.}} 
The third category consists of systems such as LlamaRL~\cite{llama_rl}, Laminar~\cite{laminar}, APRIL~\cite{april}, and SortedRL~\cite{sorted_rl}. These systems do not introduce explicit mechanisms to control or bound data staleness during execution, allowing rollout and training to proceed fully asynchronously without constraints.

\mysubsubsection{Discussions} \reviewerA{Building on the results in Figure~\ref{fig:different_staleness}, we draw the following conclusions. 
First, systems without staleness guarantees, while potentially enabling higher system throughput, may incur an unbounded \textit{staleness bound} $\eta$ that can ultimately harm convergence. 
Second, systems with one-step staleness provide a bounded staleness guarantee but forgo the opportunity to relax $\eta$ (i.e., allowing $\eta \ge 2$) in pursuit of further performance gains. 
Therefore, the approach with strict staleness control is the most favorable, as it offers significant opportunities for performance optimization while maintaining a bounded staleness guarantee.}




\begin{figure}[!t]
    \centering
    \includegraphics[width=\linewidth]{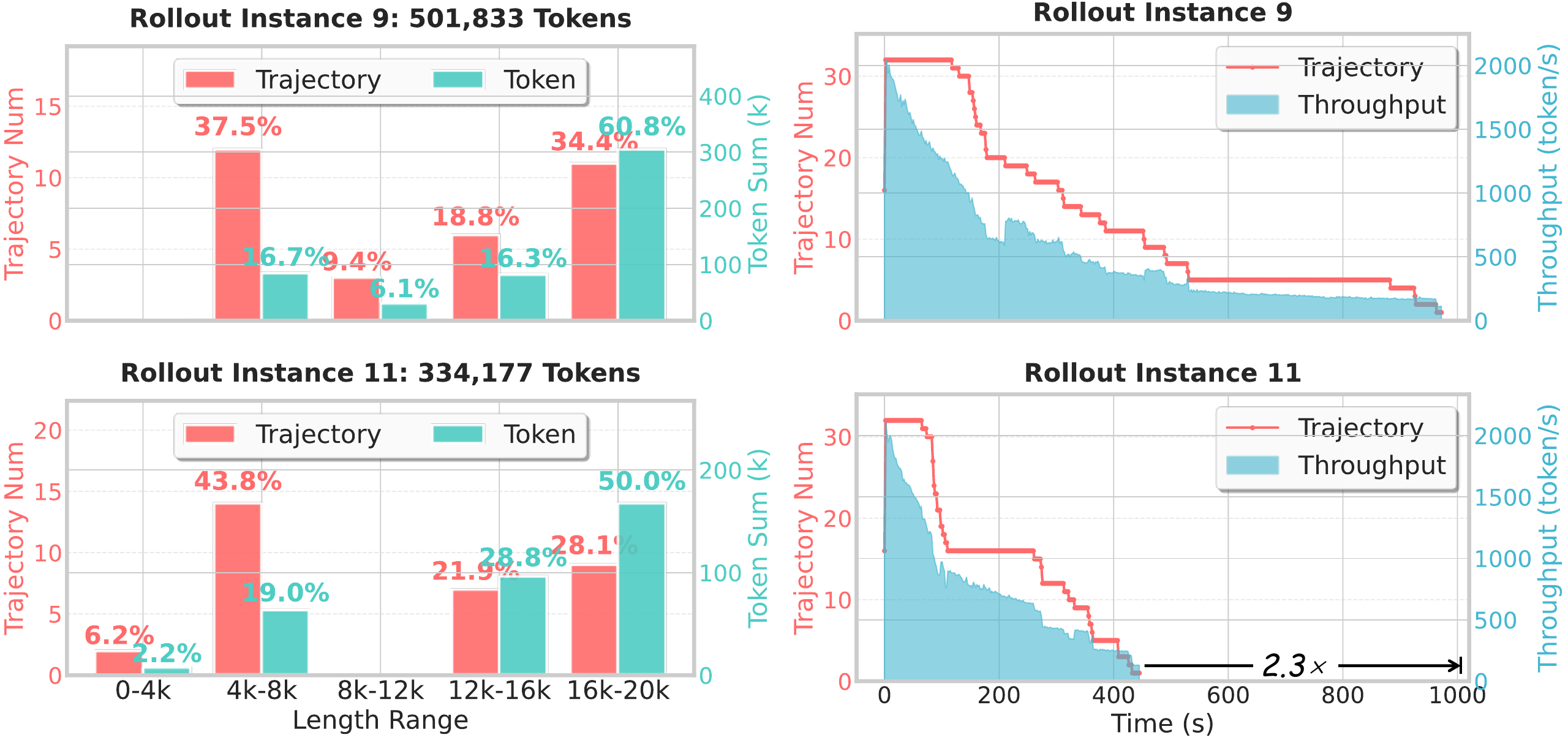}
    \myvspace{-18pt}
    \caption{\small{A rollout step of Qwen3-30B-A3B (128 H20 GPUs). (Left) Trajectory and token distributions are highly skewed both \textit{within} each instance and \textit{across} different instances. (Right) This causes intra-instance underutilization and inter-instance idle waiting.}}
    \label{fig:skewness}
    \myvspace{-5pt}
\end{figure}

\subsection{Data Skewness in RL}
\label{subsec:data_skewness}


\mysubsubsection{Impacts}
\reviewerA{Data staleness arises from model version mismatches during asynchronous rollout generation and training. In contrast, data skewness is inherent to RL execution. As illustrated in Figure~\ref{fig:skewness} (left), the auto-regressive nature of large models leads to variable-length rollout trajectories. This variability induces the issue shown in Figure~\ref{fig:skewness} (right):} First, \textit{within an instance}, shorter trajectories finish quickly, leaving the instance occupied for a prolonged period by a small number of long-tail trajectories, thus keeping throughput low. Second, \textit{across instances}, some instances finish much faster than others; faster instances therefore must wait for the slowest one before the system can synchronize model parameters globally. Both types of skewness constrain the overall system throughput.

\begin{figure}[!t]
    \centering
    \includegraphics[width=\linewidth]{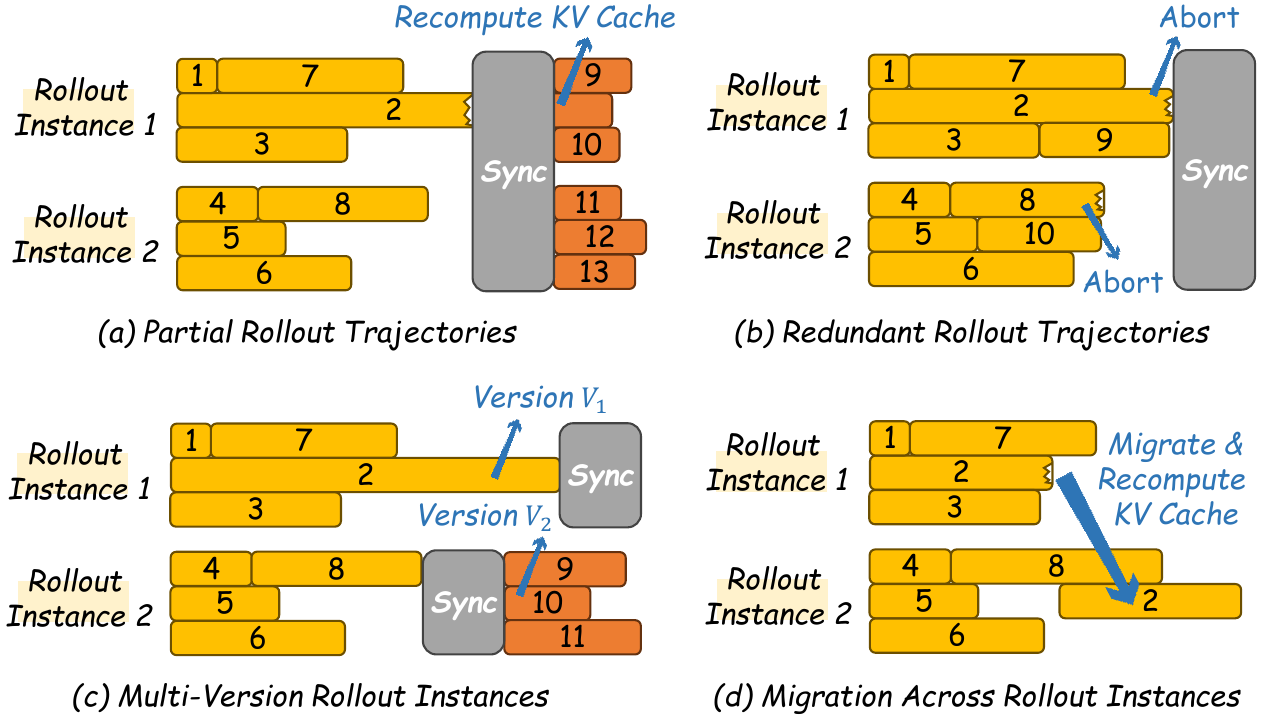}
    \myvspace{-20pt}
    \caption{\small{Rollout coordination techniques for mitigating skewness.}}
    \label{fig:techniques}
    \myvspace{-5pt}
\end{figure}

\mysubsubsection{Current solutions}
To mitigate these issues, several rollout coordination techniques have been proposed, as summarized in Figure~\ref{fig:techniques}.

(1) \textit{\textbf{Partial rollout trajectories~\cite{llama_rl, ulorl, areal, kimi_partial_rollout, roll_flash, asyncflow}.}}
First, model synchronization does not need to wait for the rollout to fully complete.
As shown in Figure~\ref{fig:techniques}(a), we can interrupt an ongoing trajectory, synchronize the model, recompute the KV Cache (i.e., re‑prefill), and then resume generation. This allows the instance to promptly admit data from newer versions (e.g., trajectories~9 and~10), thereby alleviating \textit{within-instance} skewness.

(2) \textit{\textbf{Redundant rollout trajectories~\cite{april, sorted_rl, roll_packer}.}}
Another technique is to oversubscribe the number of trajectories. For example, in Figure~\ref{fig:techniques}(b), once the required number of trajectories (e.g., 8) have been generated, the remaining ones (e.g., trajectories~2 and~8) can be aborted. This naturally filters out long‑tail trajectories and mitigates \textit{within-instance} skewness.

(3) \textit{\textbf{Multi-Version rollout instances~\cite{laminar, rhy_rl, asyncflow}.}}
While the methods above alleviate \textit{within‑instance} skewness, faster instances may still be blocked by slower ones. To address this, as shown in Figure~\ref{fig:techniques}(c), different instances can independently synchronize the model. Instances running at different speeds may thus generate trajectories using different model versions, eliminating the need for global synchronization and reducing \textit{across-instance} skewness.

(4) \textit{\textbf{Migration across rollout instances~\cite{seer, laminar, roll_packer, rhy_rl}.}}
A further technique involves migrating trajectories across rollout instances. As illustrated in Figure~\ref{fig:techniques}(d), a running trajectory (e.g., trajectory~2) can be interrupted and migrated to another instance, where it is re‑prefilled and generation resumes. This approach mitigates both  \textit{within-instance} skewness (by migrating long-tail trajectories) and \textit{across-instance} skewness (by balancing load across faster and slower instances).


\mysubsubsection{Discussions} 
\reviewerA{Although diverse rollout coordination techniques have been proposed, existing systems usually incorporate them separately and in an ad hoc manner. For example, VeRL-Async~\cite{verl, verl_async_doc} relies on global synchronization and only considers partial rollout.  Laminar~\cite{asyncflow} supports per-instance model synchronization (i.e., multi-version rollout), but only considers rollout migration during synchronization (it does not consider partial rollout during synchronization or rollout migration in other cases). This leaves an important open question: how to systematically unify these techniques and invoke them at appropriate times.}

\subsection{\reviewerA{Challenges and Our Solutions}}
\label{subsec:motivation}

\reviewerA{
As discussed in \S\ref{subsec:data_staleness} and \S\ref{subsec:data_skewness}, an ideal system for addressing data staleness and skewness should simultaneously achieve two goals: (1) allowing users to specify arbitrary $\eta$ and enforcing strict staleness control, (2) unifying diverse rollout coordination techniques and triggering them at appropriate times for performance optimization. However, achieving these goals faces two key challenges.
}

\reviewerA{
(C1) \textit{\textbf{How to design a more fine-grained staleness control mechanism.}}
As illustrated in Figure~\ref{fig:compare_sys}, existing systems face an inherent trade-off between data skewness mitigation and data staleness
control. Approaches that aggressively mitigate data skewness tend to break down staleness control, while enforcing strict staleness guarantees constrains rollout coordination flexibility, making skewness difficult to handle. The root cause lies in the coarse granularity of current staleness control mechanisms (see \S\ref{subsec:data_staleness}), which assume a single model version on the rollout side and merely track the amount of in-flight data. Such a design falls short when more flexible rollout coordination is needed: multi-version rollout instances violate the single-version assumption, and redundant rollout can exceed the predefined in-flight data capacity. Consequently, designing a more fine-grained staleness control mechanism that enables flexible rollout coordination for skewness mitigation, while preserving staleness guarantees, remains an unresolved challenge.
}

\reviewerA{
(C2) \textit{\textbf{How to unify diverse rollout coordination techniques.}}
As shown in Figure~\ref{fig:compare_sys}, existing systems provide fragmented support for rollout coordination techniques, with each technique designed and optimized largely in isolation. From a systems perspective, different coordination techniques introduce complex management requirements for trajectories and model parameters, yet existing systems lack a unified abstraction to coordinate them efficiently. From an algorithmic perspective, existing systems largely optimize each technique independently, missing opportunities for global optimization across techniques. As a result, jointly leveraging diverse coordination techniques in a unified and adaptive manner remains an open challenge.
}

\reviewerA{
To this end, we present \system. To address (C1), we redesign staleness control at the trajectory level, since rollout coordination is fundamentally performed over trajectories (\S\ref{subsec:version_identifier}). Unlike prior coarse-grained designs that track only the amount of in-flight data, \system introduces a novel consistency protocol that tracks the lifecycle of every trajectory (\S\ref{subsec:protocol}), thereby supporting flexible rollout coordination (\S\ref{subsec:staleness_advance}). To address (C2), at the system level, we introduce two data servers (i.e., a trajectory server and a parameter server) as middleware, which enable unified and flexible data movement during rollout coordination (\S\ref{subsec:arch}). At the algorithmic level, we build on a snapshot-command cycle (\S\ref{subsec:cycle}) to design a series of throughput-oriented strategies that implicitly subsume all rollout coordination techniques, as well as their combinations, and automatically invoke the coordination at appropriate times (\S\ref{subsec:strategy})
}

\reviewerA{
Together, these designs enable \system to mitigate data skewness in a unified and flexible manner, maximizing system performance while preserving strict staleness control.
}

%% file: sections/overview.tex
\section{Overview}
\label{sec:overview}

\begin{figure}[!t]
\centering
\includegraphics[width=\linewidth]{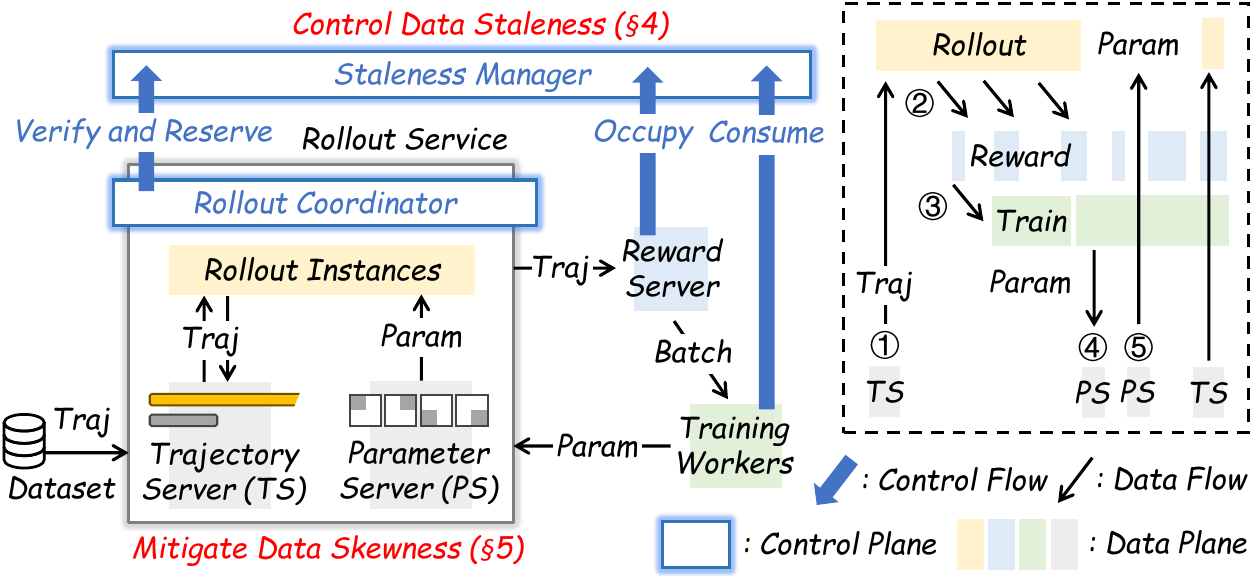}
\myvspace{-20pt}
\caption{\small{(Left) Overview of \system. A staleness manager enforces strict data staleness control, while a rollout coordinator and two data servers provide flexible, efficient rollout coordination to mitigate data skewness. (Right) Standard data flow in \system. Trajectories are sourced from the TS (\textcircled{\scriptsize 1}) and flow through the rollout, reward, and training phases (\textcircled{\scriptsize 2}–\textcircled{\scriptsize 3}). After training a batch, updated model parameters are pushed to the PS (\textcircled{\scriptsize 4}), and rollout instances selectively pull fresh parameters as needed (\textcircled{\scriptsize 5}).}}
\label{fig:overview}
\myvspace{-5pt}
\end{figure}

In this section, we present the overall design and system overview of \system, along with its end-to-end data flow. As shown in Figure~\ref{fig:overview}, our system comprises four major components: a staleness manager, a rollout service with a centralized coordinator, a reward server, and training workers. Particularly, the first two components are newly designed in our system.

\reviewerA{To enable trajectory-level data staleness control (\S\ref{sec:control_staleness}), we introduce a staleness manager that enforces a global consistency protocol. This manager continuously interacts with all system components to ensure that $\eta$ is strictly respected.} Specifically, (1) The rollout coordinator first consults the staleness manager to verify whether generating a trajectory on a given rollout instance would violate $\eta$. If admitted, rollout generation starts, and the protocol performs \texttt{Reserve} to mark the trajectory as in progress. (2) Upon completion of the rollout and reward computation, the protocol performs \texttt{Occupy} to register the trajectory as ready for consumption. (3) Once a full training batch is accumulated, the training workers invoke \texttt{Consume} to retire the batch. Through these interactions, the protocol explicitly tracks the lifecycle of each trajectory and guarantees strict adherence to the global staleness constraint.

\reviewerA{To enable unified and flexible mitigation of data skewness (\S\ref{sec:mitigate_skewness}),} we further augment the rollout service with two dedicated data servers: a trajectory server (TS) and a parameter server (PS). The TS manages all trajectories involved in rollout and mediates access between the dataset and the rollout instances, while the PS maintains the latest model parameters and provides a synchronization interface between the training workers and the rollout instances. As shown in Figure~\ref{fig:overview} (right), these servers operate as middleware that continuously supply fresh trajectories and up-to-date model parameters for rollout. The key insight is that decoupling data movement (i.e., trajectories and parameters) from rollout execution enables fine-grained control over trajectory routing and parameter synchronization. Further governed by a centralized coordinator, the rollout service thus achieves unified and flexible rollout coordination.

\begin{figure*}[!t]
    \centering
    \includegraphics[width=\linewidth]{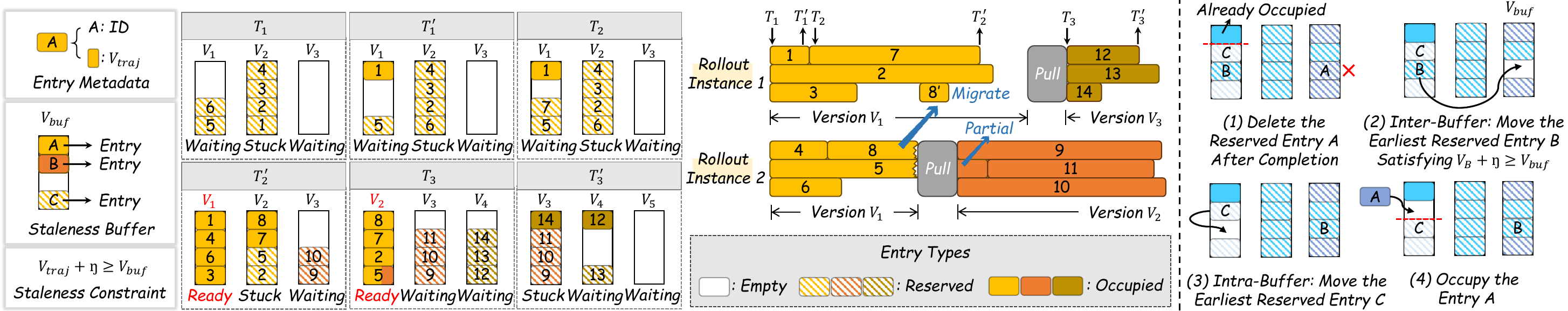}
    \myvspace{-20pt}
    \caption{\small{(Left) 
    \reviewerA{Illustration of staleness buffers. 
    A colored box (e.g., the yellow box labeled A) represents a trajectory (i.e., a buffer entry), with its color indicating $V_{traj}$. The figure shows how the same set of buffers evolves over time, illustrating how trajectories with different $V_{traj}$ are gradually reserved and occupied without violating $\eta$.}
    In this example, $\eta = 1$ (i.e., $V_{traj} + 1 \ge V_{buf}$), and each rollout instance supports up to three concurrent trajectories. (Right) Buffer transitions when a \texttt{Reserve} completes and a corresponding \texttt{Occupy} must be applied.}}
    \label{fig:staleness_control}
    \myvspace{-5pt}
\end{figure*}

%% file: sections/staleness.tex
\section{Control Data Staleness}
\label{sec:control_staleness}

This section presents how the staleness manager enforces a strict \textit{staleness bound}~$\eta$ via a global consistency protocol built around trajectory version identifiers (\S\ref{subsec:version_identifier}). The protocol is realized through a set of buffer-based primitives: \texttt{Reserve}, \texttt{Occupy}, and \texttt{Consume} (\S\ref{subsec:protocol}), and is designed to remain compatible with a wide range of \reviewerA{flexible} rollout coordination techniques (\S\ref{subsec:staleness_advance}).

\subsection{Trajectory Version Identifier}
\label{subsec:version_identifier}

Enabling flexible rollout coordination requires the ability to track data and bound staleness at the trajectory level. To achieve this, each trajectory in \system is assigned a version identifier, denoted as $V_{traj}$, which indicates the model version that will be used to generate the trajectory. \reviewerA{When} partial rollout is enabled (i.e., different segments may be generated using different versions), it refers to the oldest tolerated version.\footnote{\reviewerA{Partial rollout is enabled by default (\system works both with and without it).}} 
The assignment of $V_{traj}$ is handled jointly by the rollout coordinator and the staleness manager.

The rollout coordinator dynamically proposes $V_{traj}$ for initial trajectories before generation. 
The staleness manager, on the other hand, serves a dual role as both a discriminator and a tracker. (1) \textit{As a discriminator,} it verifies whether a proposed $V_{traj}$ would violate the specified \textit{staleness bound}~$\eta$. If not, the rollout coordinator may proceed to assign this $V_{traj}$. (2) \textit{As a tracker,} it persistently records $V_{traj}$ once assigned and maintains the lifecycle of each trajectory.

Through these two functions, the staleness manager ensures that all trajectories are assigned a $V_{traj}$, are properly tracked, and strictly adhere to the global staleness constraint.


\subsection{Global Consistency Protocol}
\label{subsec:protocol}

\mysubsubsection{Staleness buffer}
Within the staleness manager, we introduce the virtual staleness buffer abstraction, which serves as the core protocol for recording $V_{traj}$, tracking trajectories, and bounding staleness. As shown in Figure~\ref{fig:staleness_control} (left), \reviewerA{each buffer corresponds to a training batch and is associated with a $V_{buf}$, indicating the model version at which the buffer is consumed for training. The buffer consists of multiple entries, each representing a trajectory and storing its metadata, including a unique ID and its $V_{traj}$. Under this abstraction, enforcing strict staleness control reduces to placing each trajectory into an appropriate buffer such that $V_{traj}$ lies within $\eta$ versions of $V_{buf}$, i.e., satisfying $V_{traj} + \eta \ge V_{buf}$. As rollout trajectories are produced and placed into buffers, training workers continuously \texttt{Consume} staleness buffers to update the model.
}


\mysubsubsection{As a tracker} 
\reviewerA{To determine the appropriate placement for a trajectory with $V_{traj}$, we adopt a ``\texttt{Reserve} then \texttt{Occupy}'' mechanism:}
(1) \texttt{Reserve}, which places a temporary entry as a placeholder for a trajectory that has started but not yet completed. \reviewerA{As illustrated in Figure~\ref{fig:staleness_control} (left), at time $T_1$, trajectories 1–6 are assigned $V_{traj}=V_1$ and are about to be generated. Since their completion times are unknown, we \texttt{Reserve} their worst-case admissible positions (beyond which $\eta$ would be violated), corresponding to $V_{buf}=V_2$ in this example (i.e., $V_{buf}=V_{traj}+\eta$). This reservation indicates that 6 trajectories are in progress and must eventually be placed no later than $V_{buf}=V_2$.}
(2) \texttt{Occupy}, which finalizes and records a completed, rewarded trajectory in the buffer. \reviewerA{Continuing the example, at time $T_1'$, trajectory 1 completes. We release its reserved placeholder and greedily assign it to the earliest available position via \texttt{Occupy}. This determines its final placement and enables the earliest possible training.}
Through the worst-case \texttt{Reserve} and greedy \texttt{Occupy}, we can ensure the occupied position
is always no later than the reserved one, so $\eta$ is never violated.





\mysubsubsection{As a discriminator} 
When the rollout coordinator consults the protocol to verify whether a particular $V_{traj}$ can be assigned, the protocol only needs to simulate a \texttt{Reserve}: If an empty entry can still be successfully claimed, the assignment is permitted. Otherwise, a larger $V_{traj}$ is needed to unlock newer buffers for reservation, guaranteeing that the \textit{staleness bound} $\eta$ remains satisfied.



\mysubsubsection{Buffer states}
As illustrated in Figure~\ref{fig:staleness_control} (left), a staleness buffer can be in one of three states:
(1) \textit{Waiting}. The buffer still contains empty entries and \texttt{Reserve} may continue. For example, at $T_2$, buffers with $V_{buf}=V_1$ through $V_{buf}=V_2$ remain in the \textit{Waiting} state, allowing trajectory~7 to continue being assigned $V_{traj}=V_1$ and routed to instance~1. (2) \textit{Ready}. All entries are occupied, allowing the buffer to be consumed by training. (3) \textit{Stuck}. The buffer is full but contains at least one reserved (unfinished) entry. In this state, training and further reservations are blocked until the in-flight trajectories are complete. For example, at time $T'_2$, buffer $V_{buf}=V_2$ is \textit{Stuck}, preventing additional trajectories from being routed to instance~1 with $V_{traj}=V_1$, whereas buffer $V_{buf}=V_3$ is not \textit{Stuck}, allowing trajectories to be assigned $V_{traj}=V_2$ and routed to instance~2.

\mysubsubsection{Entry deletion and movement}
Before a completed trajectory can be occupied, its reserved entry must be deleted. This deletion may trigger movement of other reserved entries to ensure that \texttt{Occupy} can occur as early as possible. Figure~\ref{fig:staleness_control} (right) illustrates the procedure:
(1) Let a reserved entry A with version $V_A$ be located in buffer $V_{buf}$ and be ready for deletion.
(2) We scan buffers from smallest up to $V_{buf}$ to find the earliest reserved entry B satisfying $V_B + \eta \ge V_{buf}$, and move B to A’s position.
(3) If B’s original buffer then exposes a smaller reserved entry C, C is moved into B’s former position.
(4) Finally, \texttt{Occupy} is executed for trajectory A at the earliest empty entry. This transition keeps ready‑to‑use (occupied) entries in earlier buffers while pushing reserved entries toward later buffers, maintaining maximal readiness for training.

\mysubsubsection{Discussions}
This global consistency protocol is lightweight, as it only needs to record the metadata of trajectories (ID and $V_{traj}$) without storing the actual payload. Furthermore, the total amount of data it needs to track corresponds to the amount of in-flight data (after \texttt{Consume}, data is no longer tracked), which is at most $(\eta+1) \times \text{batch\_size}$. This depends solely on the configuration of the RL algorithm and is independent of the cluster scale.

\reviewerB{Furthermore, interactions with the protocol (e.g., \texttt{Reserve} and \texttt{Occupy}) are implemented efficiently as RPCs using Ray{~\cite{ray}}, and their worst-case time complexity is only $O(\eta+\text{batch\_size})$. Specifically, we maintain three variables for each buffer: the pointer to the last occupied entry, the pointer to the last non-reserved entry, and the smallest $V_{traj}$ in the buffer. (1) For \texttt{Reserve}, scanning backward to locate the last non-reserved entry incurs a worst-case cost of $O(\eta)$. (2) For \texttt{Occupy}, the worst-case cost includes (see Figure{~\ref{fig:staleness_control}}, right): $O(1)$ for deletion, $O(\eta + \text{batch\_size})$ for inter-buffer scanning based on the smallest $V_{traj}$, $O(\text{batch\_size})$ for intra-buffer scanning, and $O(\eta)$ for the final occupation scan. After \texttt{Reserve}/\texttt{Occupy}, updating the three variables of the modified buffer incurs an additional cost of $O(\text{batch\_size})$.}
\reviewerB{As measured in \S{\ref{subsec:interpretation}}, the protocol adds less than 1\% end-to-end latency, confirming its low overhead.}

\begin{figure}[!t]
    \centering
    \includegraphics[width=\linewidth]{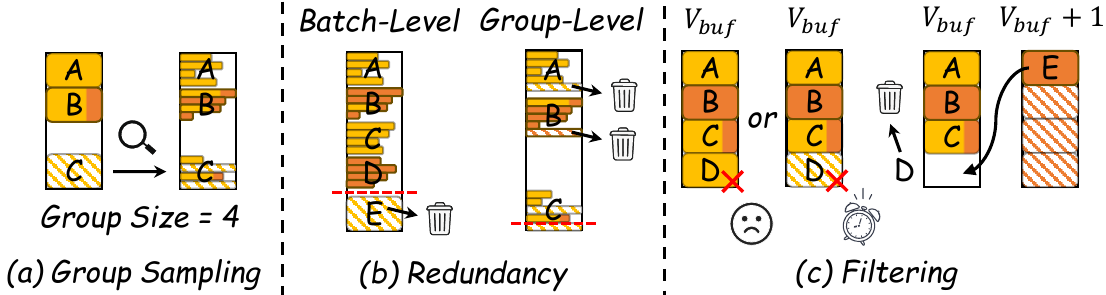}
    \myvspace{-20pt}
    \caption{\small{The staleness buffer is designed to be fully compatible with a wide range of \reviewerA{flexible} rollout coordination techniques.}}
    \label{fig:staleness_advance}
    \myvspace{-5pt}
\end{figure}

\begin{figure*}[!t]
    \centering
    \includegraphics[width=\linewidth]{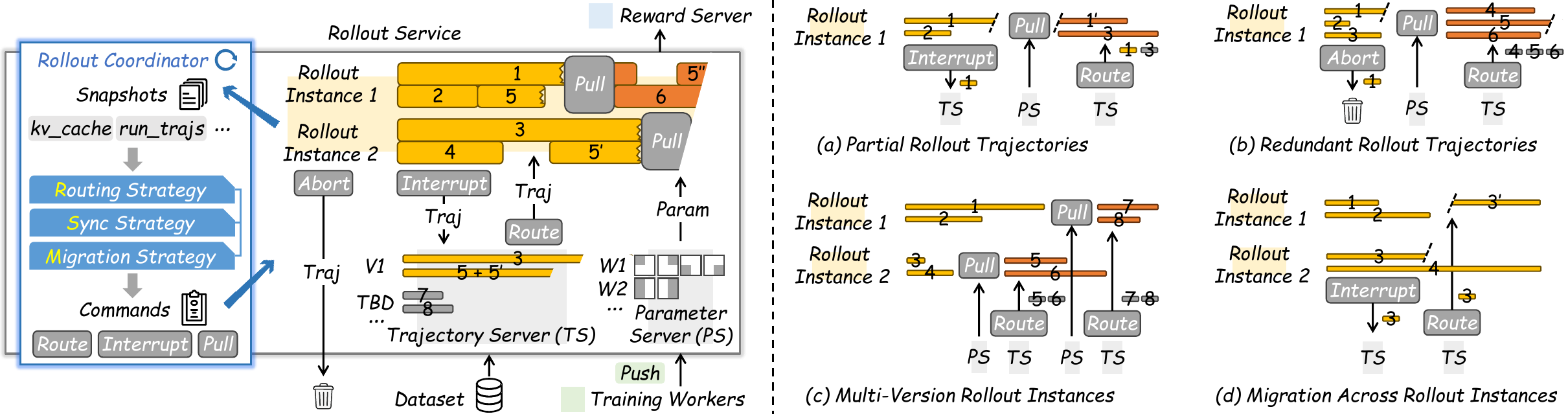}
    \myvspace{-20pt}
    \caption{\small{(Left) Rollout service architecture. The centralized coordinator periodically captures snapshots and issues commands to coordinate rollout instances, TS, and PS. (Right) Diverse rollout coordination techniques are supported via different commands based on TS and PS.}}
    \label{fig:rollout_service}
    \myvspace{-5pt}
\end{figure*}


\subsection{\reviewerA{Support for Flexible Rollout Coordination}} 
\label{subsec:staleness_advance}

\mysubsubsection{Partial rollout and rollout migration}
As illustrated in Figure~\ref{fig:staleness_control} (left), the protocol naturally supports both partial rollout and rollout migration. \reviewerA{This is because it relies only on $V_{traj}$ and is agnostic to how individual segments of a trajectory are produced, whether different segments are generated using different model versions (partial rollout) or by different rollout instances (rollout migration).} \reviewerC{The only requirement is that all segments be generated at or above the pre-assigned $V_{traj}$. Guaranteeing this requirement, as well as deciding when to trigger partial rollout and rollout migration, is handled by the rollout coordinator and its strategies (see \S\ref{subsec:strategy}).}

\mysubsubsection{Group sampling}
RL algorithms may specify a group size alongside the batch size to generate multiple trajectories for the same prompt and process them as a whole~\cite{grpo,dapo,gspo}. As shown in Figure~\ref{fig:staleness_advance}(a), in such cases, we maintain buffer entries at the granularity of trajectory groups and perform \texttt{Reserve}/\texttt{Occupy} per group. A reserved entry can be deleted and occupied only when all trajectories in the group are complete (e.g., entry~C in the figure remains reserved because two trajectories in its group are still unfinished). Besides, the version recorded for the group is the minimum $V_{traj}$ across all trajectories in the group $\mathcal{G}$, i.e., $\min_{traj \in \mathcal{G}} \{V_{traj}\}$, indicating the maximum staleness tolerated for the whole group.

\mysubsubsection{Redundancy and filtering}
The staleness buffer also accommodates flexible redundant rollout and rollout filtering. (1) As shown in Figure~\ref{fig:staleness_advance}(b), redundancy can be applied at either the batch level (expanding the number of buffer entries) or the group level (increasing the number of trajectories per entry). When a buffer reaches its predefined batch size or an entry reaches its group size, surplus data can be aborted, dropping trajectories that take too long to generate. (2) As shown in Figure~\ref{fig:staleness_advance}(c), while redundancy results in passive abortion, we can also proactively filter certain occupied or reserved entries. This is beneficial when trajectories are not needed by the algorithm (e.g., identical rewards within a group provide no learning signal~\cite{dapo}) or when training is stalled waiting for excessively delayed trajectories \reviewerB{(e.g., trajectories that run unusually long or become stuck due to partial failures)}. \reviewerC{Specifically for the latter case, we record the elapsed time once a buffer becomes \textit{Stuck}. If this duration exceeds a configurable threshold (60s by default), we abort the remaining reserved entries.} After abortion, occupied entries from later buffers can be moved forward to fill the empty slot, allowing the buffer to become \textit{Ready} quickly without waiting for new trajectories to complete. \reviewerC{Notably, aborted prompts are retained in the dataset and will be resampled in subsequent epochs.}

%% file: sections/skewness.tex
\section{Mitigate Data Skewness}
\label{sec:mitigate_skewness}

\begin{figure}[!t]
    \centering
    \includegraphics[width=\linewidth]{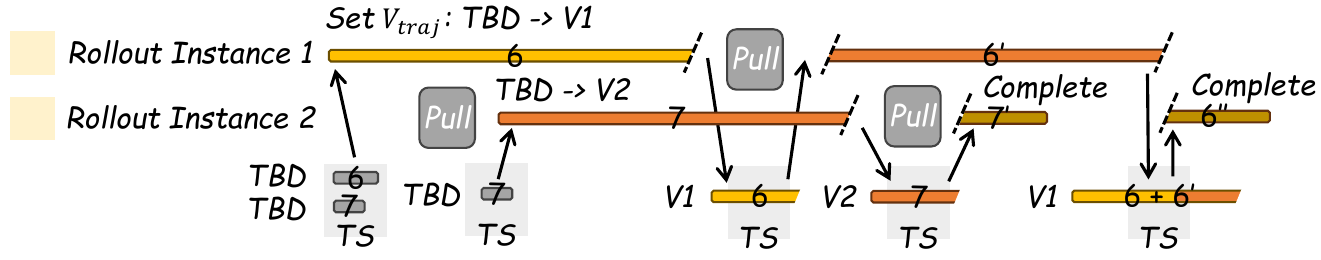}
    \myvspace{-20pt}
    \caption{\small{\reviewerA{Illustration of the trajectory server (TS).} Trajectories flow through different rollout instances via the TS as an intermediary.}}
    \label{fig:ts}
    \myvspace{-5pt}
\end{figure}

While data staleness is rigorously enforced by the global staleness manager, data skewness is handled within the rollout service through a centralized coordinator and two dedicated data servers (\S\ref{subsec:arch}). The coordinator continuously executes a snapshot-command cycle (\S\ref{subsec:cycle}) to deploy a suite of rollout coordination strategies (\S\ref{subsec:strategy}), which mitigate data skewness and enhance system performance.

\subsection{Rollout Service Architecture}
\label{subsec:arch}

\mysubsubsection{Design intuitions} Most existing RL systems manage data movement in a closed-loop manner, where trajectories are fed directly from datasets to rollout instances, and model parameters are synchronized directly from training workers to rollout instances. While straightforward, this tight coupling makes coordination on rollout instances challenging: trajectories may require interruption, resumption, or migration across instances, and breaking the global synchronization necessitates instances to update models independently. To address this, we introduce a trajectory server (TS) and a parameter server (PS) that serve as middleware storage, thereby decoupling the data movement. As illustrated in Figure~\ref{fig:rollout_service} (left), with the rollout coordinator acting as the central control plane, we can flexibly coordinate rollout instances through the support of TS and PS, enabling controlled delivery of trajectories and parameters.

\begin{figure}[!t]
    \centering
    \includegraphics[width=\linewidth]{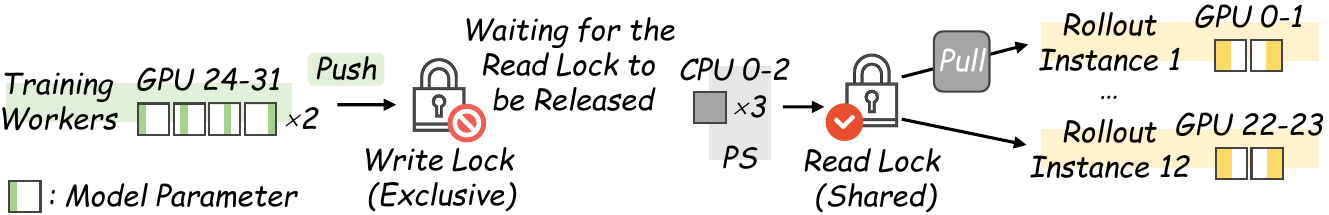}
    \myvspace{-20pt}
    \caption{\small{\reviewerA{Illustration of the parameter server (PS).} Training workers continuously \texttt{Push} updated model parameters to the PS, while rollout instances \texttt{Pull} them on demand. A read-write locking scheme is used to ensure correctness.}}
    \label{fig:ps}
    \myvspace{-5pt}
\end{figure}

\mysubsubsection{Rollout coordinator}
As shown in Figure~\ref{fig:rollout_service} (left), the coordinator periodically captures snapshots of rollout instances, applies a suite of rollout coordination  strategies based on them, and subsequently issues rollout commands. These commands include (1)~\texttt{Pull}, which instructs a designated rollout instance to fetch the latest model parameters from the PS;
(2)~\texttt{Route}, which selects a trajectory from the TS and assigns it to a specific rollout instance for generation; and
(3)~\texttt{Interrupt}, which terminates an ongoing trajectory on a given instance and returns the partially generated trajectory to the TS.\footnote{In addition, the staleness manager may directly issue \texttt{Abort} commands to rollout instances to irrevocably discard a trajectory, as discussed in \S\ref{subsec:staleness_advance}. In this case, the trajectory will not be returned to the TS.} As shown in Figure~\ref{fig:rollout_service} (right), by composing these commands and leveraging the TS and PS as middleware, rollout instances can realize a wide range of rollout coordination  techniques.


\mysubsubsection{Trajectory server}
The TS \reviewerA{is a centralized CPU server connected via a local network that} stores all trajectories required by the rollout instances. \reviewerA{It has a} default capacity of $(\eta+1) \times \text{batch\_size}$, \reviewerA{matching the
upper bound of in-flight data to be tracked. Figure~\ref{fig:ts} illustrates its workflow: trajectories are initially sampled from the dataset and placed on the TS without an assigned $V_{traj}$. After the rollout coordinator verifies admissible version candidates (\S\ref{subsec:protocol}), it assigns an appropriate $V_{traj}$ and issues \texttt{Route} to send the trajectory to a target rollout instance (e.g., trajectory 6 is assigned $V_1$ and routed to instance 1). During generation, \texttt{Interrupt} may be issued due to partial rollout (e.g., trajectory 7) or migration (e.g., trajectory 6+6'). The TS stores such interrupted trajectories and re-routes them.}

\mysubsubsection{Parameter server}
\reviewerA{The PS is a distributed CPU-based storage that maintains the latest model parameters.} The rollout coordinator issues \texttt{Pull} commands, directing specific rollout instances to retrieve parameters from the PS, while model updates on the PS are made via \texttt{Push} operations, triggered automatically by the training workers. Once a training step completes, the training workers immediately \texttt{Push} the updated parameters to the PS, overlapping the communication with the next training step. \reviewerA{In contrast, \texttt{Pull} cannot be overlapped and is therefore more latency-critical. To optimize this path, the PS shares CPU resources with rollout workers, enabling \texttt{Pull} to leverage faster PCIe-based DMA, while \texttt{Push} utilizes RDMA over InfiniBand. Additional implementation details and communication optimizations are provided in Appendix~\ref{appendix:impl}.}

\reviewerA{Figure~\ref{fig:ps} further illustrates that,} to ensure correctness, the PS adopts a read-write locking scheme inspired by database concurrency control: A \texttt{Push} operation (analogous to an exclusive write) blocks concurrent \texttt{Pull} commands, and vice versa, while multiple \texttt{Pull} commands from different rollout instances (analogous to shared reads) may proceed concurrently.

\mysubsubsection{Discussions}
While introducing the TS and PS as middleware adds extra data movement \reviewerA{(from the dataset to the TS and from training workers to the PS), the resulting overhead is minimal: For the TS,} reading pure tokens from the dataset imposes a light load. \reviewerA{For the PS,} the training workers’ \texttt{Push} operations can be overlapped. 

\reviewerA{Moreover, although other critical data movements involved in rollout coordination cannot be overlapped, their overhead, including TS-side (\texttt{Route}/\texttt{Interrupt} trajectories)
and PS-side (\texttt{Pull} parameters),
accounts for $<$3\% of the total (see Table~\ref{tab:time_breakdown} in \S\ref{subsec:interpretation}).}

\reviewerA{Last but not least}, as the cluster scales, the overhead of the TS does not grow due to its fixed capacity \reviewerA{of $(\eta+1) \times \text{batch\_size}$}, and \reviewerA{the overhead of} both \texttt{Push} and \texttt{Pull} for the PS remains constant (see Appendix~\ref{appendix:comm_scale}). Our design therefore does not affect scalability. 

\begin{figure}[!t]
    \centering
    \includegraphics[width=\linewidth]{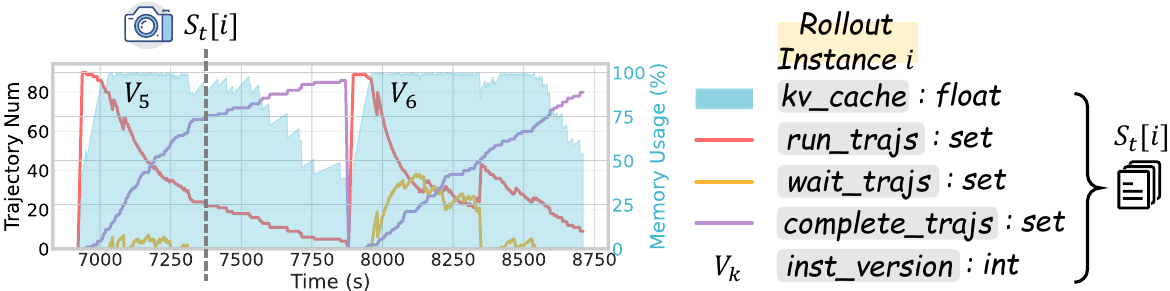}
    \myvspace{-20pt}
    \caption{\small{A snapshot captures five fields for every rollout instance.}}
    \label{fig:snapshot}
    \myvspace{-5pt}
\end{figure}

\begin{figure*}[!t]
    \centering
    \includegraphics[width=\linewidth]{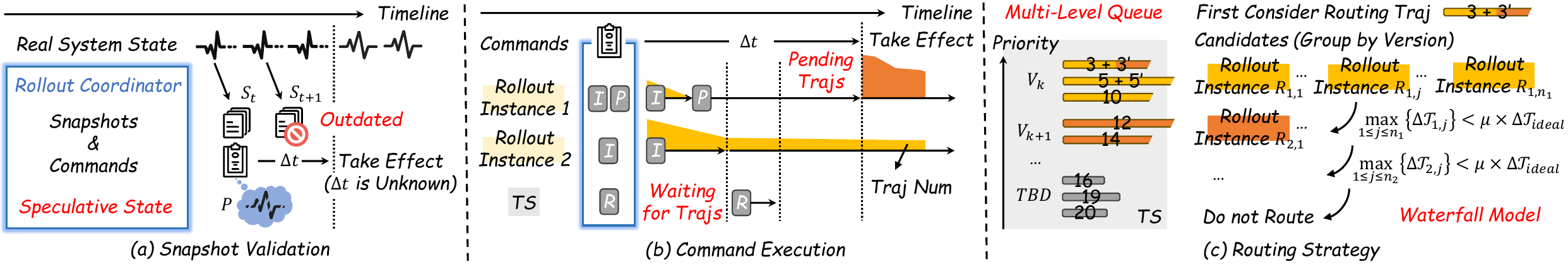}
    \myvspace{-20pt}
    \caption{\small{(a) To avoid outdated snapshots, we introduce a \textit{speculative state} ($P$) to verify whether the expected effect has occurred after issuing the commands. (b) After issuance, commands are  executed concurrently. When dependencies exist between commands, we address them by either having commands wait or keeping data pending. ``\texttt{I}'', ``\texttt{R}'' and ``\texttt{P}'' represent \texttt{Interrupt}, \texttt{Route} and \texttt{Pull}, respectively. (c) Our \textit{routing strategy} utilizes a multi-level queue (MLQ) and a waterfall model to greedily optimize the marginal throughput benefits.}}
    \label{fig:skewness_mitigation}
    \myvspace{-5pt}
\end{figure*}

\subsection{Snapshot-Command Cycle}
\label{subsec:cycle}

To mitigate data skewness, the rollout coordinator operates continuously in a snapshot-command cycle, monitoring the system’s real-time load and performing rollout coordination  strategies.

\mysubsubsection{Fields of the snapshot} Each captured snapshot $S$ aggregates five fields for every instance: the GPU memory usage of the KV Cache (\text{kv\_cache}), the trajectories currently under generation (\text{run\_trajs}), the trajectories queued in the rollout engine’s internal waiting queue (\text{wait\_trajs}), the trajectories completed on the instance since the last model synchronization (\text{complete\_trajs}), and the instance’s current model version (\text{inst\_version}). As illustrated in Figure~\ref{fig:snapshot}, during rollout, \text{run\_trajs} generally decreases while \text{complete\_trajs} monotonically increases. As trajectories grow longer, \text{kv\_cache} increases accordingly. Once it reaches the budget, trajectories from \text{run\_trajs} are preempted and moved into \text{wait\_trajs}, until \text{kv\_cache} is released. During parameter synchronization, all \text{run\_trajs} and \text{wait\_trajs} are interrupted and cleared, and \text{inst\_version} is updated to reflect the newly synchronized model.

\mysubsubsection{Snapshot validation using speculative state}
Within this cycle of periodic snapshots, decision-making, and command execution, a key challenge arises from the temporal coupling between decisions and system state changes. As illustrated in Figure~\ref{fig:skewness_mitigation}(a), when the rollout coordinator makes decisions and issues commands based on the snapshot $S_t$, these commands require a duration $\Delta t$ to take effect in the real system after issuance. If $S_{t+1}$ is captured before the effect of the commands is realized, it may reflect the system state prior to the command execution. Using such snapshots for subsequent decisions risks operating on outdated information, potentially leading to decision inconsistencies or oscillatory behavior.

Since $\Delta t$ is highly variable and context-dependent, we introduce a snapshot validation mechanism centered on a \textit{speculative state} $P$. It represents the expected system state after each issued command has fully taken effect, with two critical fields: the expected model version of each instance (inst\_version) and the expected accumulated number of trajectories (accum\_traj\_num) for each instance, including completed ones. At system startup, for every instance, both fields in $P$ are initialized to 0. As summarized in Table~\ref{tab:command}, after each command is issued, $P$ is updated correspondingly to reflect the expected effects of that command. Upon the arrival of a new snapshot $S_{t+1}$, the rollout coordinator validates it by checking, for every instance $i$, whether the following condition holds: \begin{equation}
\small
\begin{aligned}
& P[i].\text{inst\_version} = S_{t+1}[i].\text{inst\_version} \land P[i].\text{accum\_traj\_num} =  \\
&| S_{t+1}[i].\text{run\_trajs}
\cup S_{t+1}[i].\text{wait\_trajs}
\cup S_{t+1}[i].\text{complete\_trajs} |.
\end{aligned}
\end{equation}
A snapshot is accepted only if the condition is satisfied, indicating that all previously issued commands have taken effect; otherwise, it is discarded and the coordinator waits for the next snapshot.

\begin{table}[!t]
\centering
\caption{\small{Command arguments and their effects on the speculative state ($P$). We denote ``inst'' as the targeted instance ID, ``trajs'' as the set of trajectory IDs to manipulate, and $|\cdot|$ as set cardinality.}}
\myvspace{-10pt}
\label{tab:command}
\small
\begin{tabular}{c|c|c}
\hline\toprule
Commands & Arguments & Effects on $P$ After Issuance \\
\midrule
\specialcell{\texttt{Pull}} & 
\specialcell{inst}
& \specialcell{$P[\text{inst}].\text{inst\_version} = \text{get\_ps\_version()}$\\$P[\text{inst}].\text{accum\_traj\_num} = 0$} \\
\hline
\specialcell{\texttt{Route}} & 
\specialcell{inst, trajs}
& \specialcell{$P[\text{inst}].\text{accum\_traj\_num} \mathrel{+}= |\text{trajs}|$} \\
\hline
\specialcell{\texttt{Interrupt}} & 
\specialcell{inst, trajs}
& \specialcell{$P[\text{inst}].\text{accum\_traj\_num} \mathrel{-}= |\text{trajs}|$} \\
\hline
\specialcell{\texttt{Abort}} &
\specialcell{inst, trajs}
& \specialcell{$P[\text{inst}].\text{accum\_traj\_num} \mathrel{-}= |\text{trajs}|$} \\
\bottomrule\hline
\end{tabular}
\myvspace{-10pt}
\end{table}

\begin{algorithm}[t]
\caption{Rollout coordination procedure (detailed pseudocode for each strategy is given in Appendix~\ref{appendix:alg}).}
\label{alg:overall}
\small
\KwIn{$S$: the snapshot of all rollout instances (already validated)}
$\text{ts\_trajs} \gets \text{get\_ts\_trajs}()$ \;
$\text{ps\_version} \gets \text{get\_ps\_version}()$ \;

\ForEach{$\text{inst} \in \text{SynchronizationStrategy}(S, \text{ts\_trajs}, \text{ps\_version})$}{
    $\text{trajs} \gets S[\text{inst}].\text{run\_trajs} \cup  S[\text{inst}].\text{wait\_trajs}$ \;
    $\textit{\textsc{IssueCommand}}(\texttt{Interrupt}(\text{inst}, \text{trajs}), \texttt{Pull}(\text{inst}))$ \;
    $S[\text{inst}].\text{discard}(\text{trajs})$ \;
    $S[\text{inst}].\text{inst\_version} \gets \text{ps\_version}$ \;
    $\text{ts\_trajs} \gets \text{ts\_trajs} \cup \text{trajs}$ \;
}

\ForEach{$(\text{inst}, \text{trajs}) \in \text{MigrationStrategy}(S)$}{
    $\textit{\textsc{IssueCommand}}(\texttt{Interrupt}(\text{inst}, \text{trajs}))$ \;
    $S[\text{inst}].\text{discard}(\text{trajs})$ \;
    $\text{ts\_trajs} \gets \text{ts\_trajs} \cup \text{trajs}$ \;
}

\ForEach{$(\text{inst}, \text{trajs}) \in \text{RoutingStrategy}(S, \text{ts\_trajs})$}{
    $\textit{\textsc{IssueCommand}}(\texttt{Route}(\text{inst}, \text{trajs}))$ \;
}
\end{algorithm}

\mysubsubsection{Concurrent command execution}
As shown in Figure~\ref{fig:skewness_mitigation}(b), to maximize resource utilization, commands across different data planes are executed concurrently. However, this concurrency introduces cross-command dependencies. For example, a \texttt{Route} command must wait until all interrupted trajectories have been returned to the TS. Likewise, trajectories newly routed to an instance may be delayed if that instance is currently executing a \texttt{Pull} command to update its model. To preserve execution correctness in these cases, we enforce logical ordering by either having commands wait or by keeping the data pending in the relevant data planes, thereby excluding them from execution until the dependencies are resolved.

\subsection{Rollout Coordination Strategies}
\label{subsec:strategy}

\reviewerA{As shown in Algorithm~\ref{alg:overall}, \system employs a set of strategies. Given a system snapshot, these strategies are applied sequentially to issue commands that steer rollout coordination.}

\mysubsubsection{Design intuition}
\reviewerA{A central question is how to systematically design strategies that unify diverse rollout coordination techniques. Our key observation is that existing techniques, as well as their possible combinations, can be abstracted into three fundamental behaviors: synchronization, migration, and routing.
Among them, routing serves as the unifying core, as both synchronization and migration ultimately reduce to routing decisions: synchronization interrupts trajectories and subsequently re-routes them, while migration directly re-routes trajectories across instances. As shown in Algorithm~\ref{alg:overall}, this abstraction enables us to formulate rollout coordination as a routing-centric problem, where synchronization and migration are optionally triggered as needed, and routing accounts for their effects (i.e., interrupted trajectories and model updates). Given the central role of routing, we design a precise \textit{routing strategy} that leverages a cost model to accurately estimate throughput and make optimized decisions. In contrast, we design the \textit{synchronization} and \textit{migration} strategies to be lightweight and heuristic: they only determine when and where synchronization or migration should be triggered, while deferring the remaining decision-making to the \textit{routing strategy}. In the following, we outline the core ideas of each strategy. 
Detailed pseudocode is provided in Appendix~\ref{appendix:alg}.}


\begin{figure*}[!t]
    \centering
    \includegraphics[width=\linewidth]{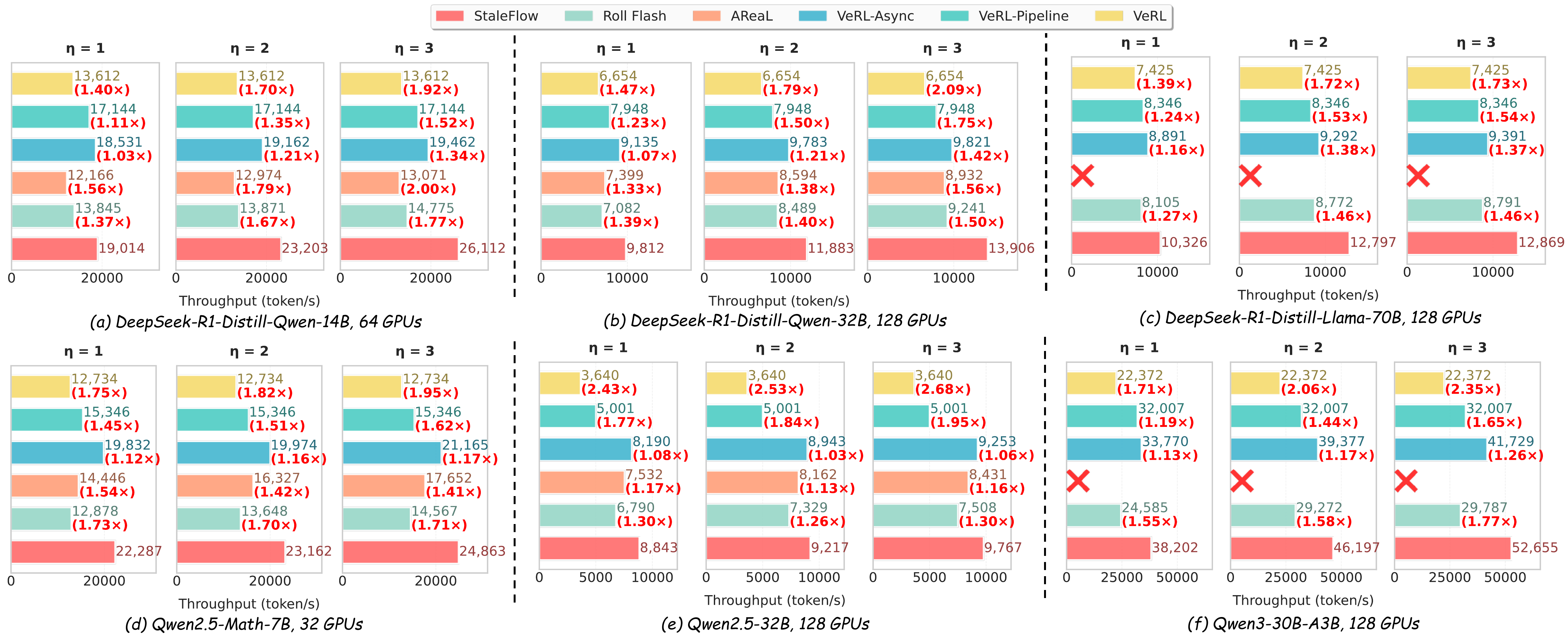}
    \myvspace{-20pt}
    \caption{\small{End-to-end throughput under different \textit{staleness bounds} ($\eta$). 
    Values in parentheses are the gains achieved by \system compared to the baselines.
    Red crosses denote unsupported configurations: data skewness causes severe load imbalance, leading to NCCL~\cite{nccl, nccl_doc} timeouts.}}
    \label{fig:e2e}
    \myvspace{-5pt}
\end{figure*}

\mysubsubsection{Cost model} 
We first develop a cost model to estimate the throughput change induced by a routing decision. Given a snapshot $S$, the generation throughput of instance $i$, denoted by $\mathcal{T}_i$, can be estimated as follows (the detailed derivation is provided in Appendix~\ref{appendix:cost_model}):
\begin{equation}
\small
\begin{aligned}
\mathcal{T}_{i}(S)= \frac{|S[i].\text{run\_trajs}|}{k_1 \times S[i].\text{kv\_cache} + \max(k_2,\; k_3 \times |S[i].\text{run\_trajs}|) + k_4}.
\end{aligned}
\end{equation}
Here, $|\cdot|$ denotes set cardinality and $k_1 \sim k_4$ are constant coefficients obtained via offline profiling and linear regression. Based on this cost model, we further derive the marginal throughput gain $\Delta \mathcal{T}_i$ when routing a trajectory $\tau$ of length $l$ to instance $i$:
\begin{equation}
\small
\begin{aligned}
& \gamma_i = \mathbb{I}(S[i].\text{kv\_cache} +k_5 
 \times l \le M \land |S[i].\text{wait\_trajs}|=0). \\
& S'[i].\text{kv\_cache} = S[i].\text{kv\_cache} +  \gamma_i\times k_5 
 \times l. \\
& S'[i].\text{run\_trajs} = S[i].\text{run\_trajs} \cup \{ \tau \mid \gamma_i = 1 \}. \\
& \implies \Delta \mathcal{T}_i = \mathcal{T}_{i}(S') - \mathcal{T}_{i}(S).
\end{aligned}
\end{equation}
In this expression, $\mathbb{I}(\cdot)$ is the indicator function, $k_5$ denotes the per-token KV Cache footprint obtained via profiling, and 
$M$ is the KV Cache budget. A routed trajectory $\tau$ can run immediately (i.e., $\gamma_i = 1$) only if it fits within the KV Cache budget and the rollout engine’s internal waiting queue is empty. Otherwise, it is placed into the waiting queue and contributes no throughput gain.

\mysubsubsection{Routing strategy} Using the above cost model, we address how to route trajectories stored in TS under a given snapshot $S$ (Algorithm~\ref{alg:overall}, line 13). As shown in Figure~\ref{fig:skewness_mitigation}(c), trajectories are first organized into a multi-level queue (MLQ) structure~\cite{mlq_1, mlq_2} ordered by increasing $V_{traj}$. Smaller $V_{traj}$ indicates higher staleness and thus higher routing priority, while initial trajectories without $V_{traj}$ are assigned the lowest priority. This ordering reflects the principle that trajectories already identified as staler should be processed earlier. Accordingly, the routing procedure proceeds from the highest-priority non-empty queue; lower-priority queues are only considered once all higher-priority ones have been exhausted.

For a trajectory of length $l$, we first identify all candidate instances where routing is admissible. An instance $i$ qualifies if either: 
\begin{itemize}[noitemsep, topsep=0pt, parsep=0pt, partopsep=0pt, leftmargin=*]
\item The trajectory is initial, and the staleness manager permits assigning $V_{traj}=S[i].\text{inst\_version}$ without violating $\eta$ (\S\ref{subsec:protocol}).
\item The trajectory already has a $V_{traj}$, indicating it is partially generated and interrupted, requiring re-routing. In this case, the re-routed instance must satisfy $S[i].\text{inst\_version} \ge V_{traj}$.
\end{itemize}
These candidate instances are then grouped by increasing inst\_version, yielding $\{\{R_{i,j}\}_{j=1}^{n_i} \}_{i=1}^{m}$. For one thing, instances with a smaller version should be prioritized, as they admit fewer feasible routing options. For another, we aim to maximize throughput by selecting the instance with the largest marginal throughput gain.

To balance priority and throughput, we leverage a simplified waterfall model~\cite{waterfall_1, waterfall_2}. We define $\Delta \mathcal{T}_{ideal}$, which denotes the marginal throughput gain achieved by routing the trajectory to an idle instance, serving as an upper bound on the achievable gain:
\begin{equation}
\small
\begin{aligned}
\Delta \mathcal{T}_{ideal} = \frac{1}{k_1 \times (0 + k_5 \times l) + \max(k_2,\; k_3 \times 1) + k_4} - 0 .
\end{aligned}
\end{equation}
As shown in Figure~\ref{fig:skewness_mitigation}(c), starting from the highest-priority instance group, the algorithm selects the instance with the largest estimated gain $\max_{1\le j\le n_1}\{\Delta \mathcal{T}_{1,j}\}$. If this gain exceeds the threshold 
$\mu 
\times \Delta \mathcal{T}_{ideal}$, the trajectory is routed to that instance; otherwise, the algorithm proceeds to the next group (i.e., $\{R_{2,1},\ldots,R_{2,n_2}\}$) and repeats the process. If no candidate instance yields a marginal gain exceeding the threshold, the trajectory is temporarily withheld from routing, allowing ongoing workloads on these instances to complete so that a higher marginal gain may be achieved in subsequent decisions.

\mysubsubsection{Synchronization strategy} We selectively synchronize model parameters to rollout instances only when synchronization is expected to improve system throughput (Algorithm~\ref{alg:overall}, line 3). (1) First, an instance $i$ is considered eligible for synchronization only if its local model version lags behind the PS, i.e., $\text{ps\_version} > S[i].\text{inst\_version}$, and no trajectory in the TS can be routed to this instance given the staleness constraint. In this case, the instance cannot accept additional load without synchronization. (2) Second, we tentatively update the instance and execute the \textit{routing strategy} once. If this tentative update enables new trajectories to be routed (indicating that the marginal gain exceeds the predefined threshold), we perform the synchronization to improve throughput.

\mysubsubsection{Migration strategy} 
During rollout, load imbalance motivates trajectory migration across instances (Algorithm~\ref{alg:overall}, line~9). We trigger migration in two cases.
(1) When the KV Cache is full, run\_trajs are preempted into wait\_trajs (Figure~\ref{fig:snapshot}), which do not contribute to throughput. We therefore set a threshold $\varphi_{wait}$: once the number of wait\_trajs on an instance exceeds this limit, the excess trajectories are interrupted and returned to the TS for possible re-routing by the \textit{routing strategy}.
(2) When the throughput gap between the highest- and lowest-throughput instances exceeds $\varphi_{throughput}$, all trajectories on the highest-throughput instance are interrupted and returned to the TS for redistribution by the \textit{routing strategy}.

%% file: sections/exp.tex
\begin{table}[!t]                                        
  \centering                                                 
  \caption{\small{Models and GPU allocations used in our evaluations.}}                   
  \myvspace{-10pt}
  \small              
  \begin{tabular}{lcc}
  \toprule
  Category & Model & GPUs \\
  \midrule
  \multirow{3}{*}{\makecell{DeepSeek-R1-Distill, Dense}} & Qwen2.5-14B & 64 GPUs \\
  & Qwen2.5-32B & 128 GPUs \\
  & \reviewerC{Llama3.1-70B} & \reviewerC{128 GPUs} \\
  \midrule
  \multirow{2}{*}{\makecell{Qwen, Dense}} & \reviewerC{Qwen2.5-Math-7B} & \reviewerC{32 GPUs} \\
  & Qwen2.5-32B & 128 GPUs \\
  \midrule
  \multirow{1}{*}{\makecell{Qwen, MoE}} & Qwen3-30B-A3B & 128 GPUs \\
  \bottomrule
  \end{tabular}
  \label{tab:models}
  \myvspace{-10pt}
\end{table}

\section{Evaluations}
\label{sec:exp}

In this section, we evaluate \system by comparing it against a broad set of baselines to assess end-to-end system performance (\S\ref{subsec:e2e}) and RL convergence (\S\ref{subsec:convergence}). We further examine its scalability (\S\ref{subsec:scalability}) and provide a detailed analysis of its performance gains (\S\ref{subsec:interpretation}).

\subsection{Experimental Setup}
\label{subsec:setup}

\mysubsubsection{Testbed and workloads}
Our experiments use a testbed of 16 machines, each with 8 NVIDIA H20 GPUs. \reviewerC{We adopt the DAPO algorithm~\cite{dapo}. For the smallest setting (i.e., Qwen2.5-Math-7B with 32 GPUs), we train and evaluate on a mixture of the GSM8K~\cite{gsm8k} and MATH~\cite{math} datasets. For the remaining larger settings, we train on the DAPO-Math-17k dataset~\cite{dapo_dataset} and evaluate on AIME24~\cite{aime24} (following prior work~\cite{laminar, roll_flash}).} The prompt and response length limits are set to 2k and 20k, respectively. The batch size is set to 128, and the group size is set to 16, meaning each training step processes a total of 2048 trajectories.

\mysubsubsection{Models and metrics}
As summarized in Table~\ref{tab:models}, we evaluate four models from the Qwen family~\cite{qwen_2, qwen_3}, covering a range of model scales and architectures. This includes both distilled~\cite{distill_survey} and base variants, as well as both dense and mixture-of-experts (MoE)~\cite{gshard, moe_survey_1, moe_survey_2} models. For each model, we allocate an appropriate cluster size to accommodate its computational requirements.

We measure system performance by reporting throughput (tokens/s), calculated as the total number of tokens processed across multiple post-training steps divided by the total execution time. To assess learning convergence, we plot reward curves during training and report the pass@1 accuracy (i.e., the proportion of problems solved correctly on the first attempt) on the evaluation set.

\mysubsubsection{Baselines and settings}
We select five baselines from three categories:  
(1) \textit{No staleness}. We first compare with VeRL~\cite{verl}, a widely used synchronous RL system where rollout and training share the same resources and execute sequentially, without introducing data staleness. 
(2) \textit{One-Step staleness}. Next, we compare with VeRL-Pipeline~\cite{verl_pipeline, verl_pipeline_doc}, which allows only one-step asynchrony. It adopts a disaggregated architecture and overlaps rollout which is one version behind with the current training step.
(3) \textit{Strict staleness control}. These baselines share the same property as \system, allowing a user-defined \textit{staleness bound} $\eta$ to tune system performance. Specifically, we compare against VeRL-Async~\cite{verl_async_doc}, AReaL~\cite{areal}, and Roll Flash~\cite{roll_flash}. Although their engineering implementations differ, they all regulate staleness by limiting the amount of in-flight data and only support partial rollout (Figure~\ref{fig:compare_sys}). 

In all experiments, we tune the GPU allocation ratio between rollout and training to optimal values and adopt the most efficient parallelism strategy. For \system, the hyperparameters related to rollout coordination  strategies are set as follows: $\mu = 0.3$, $\varphi_\text{wait} = 3$, and $\varphi_\text{throughput} = 5$. To ensure a fair comparison, redundant rollout is not used, as it would abort trajectories and negatively affect response length. Interested readers can refer to Appendix~\ref{appendix:redundant} for additional details and an ablation study on redundant rollout.

\subsection{End-to-End Throughput}
\label{subsec:e2e}

As shown in Figure~\ref{fig:e2e}, we first compare the end-to-end throughput across different systems. Under varying models and \textit{staleness bounds}, \system consistently achieves the highest throughput. (1) Compared to synchronized systems without staleness (VeRL), \system improves throughput by up to 2.68$\times$ (\reviewerC{1.91}$\times$ on average). (2) Compared to one-step staleness systems (VeRL-Pipeline), \system achieves up to 1.95$\times$ (\reviewerC{1.51}$\times$ on average) greater improvement. (3) Compared to the best-performing system among those with strict staleness control (VeRL-Async), \system still delivers a throughput gain of up to 1.42$\times$ (\reviewerC{1.18}$\times$ on average). 

These results can be explained as follows. (1) Synchronized systems cannot execute asynchronously and leverage stale data to mitigate skewness, resulting in the lowest throughput. (2) One-Step staleness systems cannot further improve performance by relaxing the \textit{staleness bound}. (3) Other systems with strict staleness control (e.g., VeRL-Async, Roll Flash, AReaL) rely on simple mechanisms that limit the number of in-flight trajectories, preventing the use of more flexible rollout coordination techniques (Figure~\ref{fig:compare_sys}). In contrast, \system supports a suite of staleness-aware rollout coordination strategies (\S\ref{subsec:strategy}) that more effectively mitigate data skewness (further evaluated in \S\ref{subsec:interpretation}). As a result, under the same \textit{staleness bound}, \system can extract substantially higher throughput gains.

\reviewerC{It can be observed that \system has a modest improvement over baselines on Qwen2.5-32B (Figure~\ref{fig:e2e}(e)). This is because it is not a reasoning model, with an average response length of only $\sim$0.5K, leading to less pronounced data skewness. In contrast, most of the other evaluated models are reasoning models with substantially longer responses ($\sim$10K), which lead to much more severe skewness. Since longer responses typically indicate stronger reasoning capability and are becoming an emerging trend, the benefits of \system are more pronounced.}

\reviewerC{Moreover, the improvement of \system further increases as the \textit{staleness bound} $\eta$ grows. At $\eta=3$, the gains over VeRL-Async become substantial (e.g., 1.34$\times$, 1.42$\times$, and 1.37$\times$ for the DeepSeek-R1-Distill model series in Figure~\ref{fig:e2e}). As tolerating larger $\eta$ is an active research direction in RL algorithms (e.g., M2PO~\cite{M2PO}, VESPO~\cite{VESPO}, and GAC~\cite{GAC}), \system’s stronger ability to mitigate data skewness under larger $\eta$ will become increasingly valuable.}

\begin{figure}[!t]
    \centering
    \includegraphics[width=\linewidth]{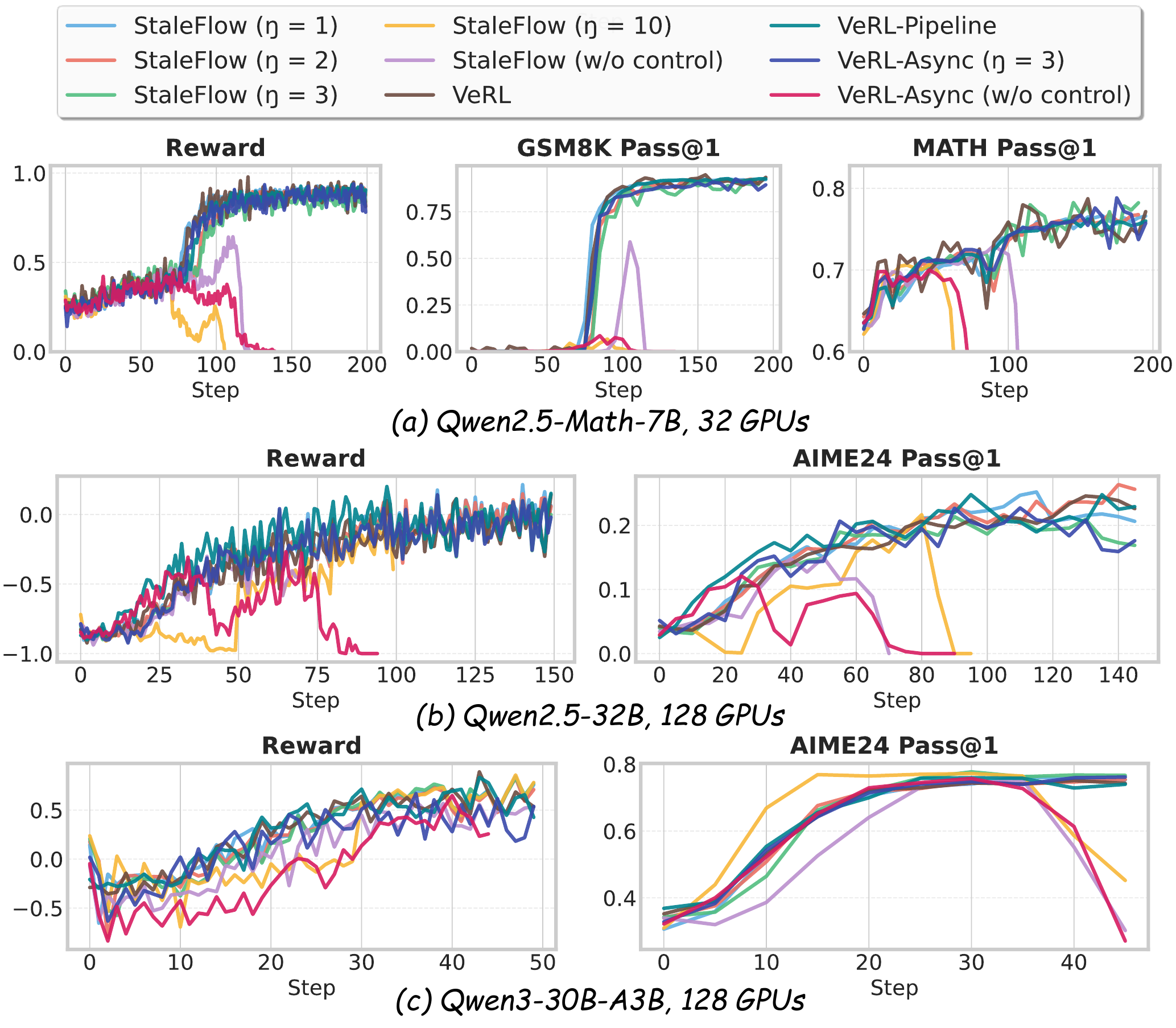}
    \myvspace{-20pt}
    \caption{\small{RL convergence comparing \system with VeRL \reviewerC{\reviewerB{and other VeRL variants. ``w/o control'' denotes the modified version with staleness control completely disabled.}}}}
    \label{fig:convergence}
    \myvspace{-5pt}
\end{figure}

\subsection{Convergence}
\label{subsec:convergence}

\reviewerC{We next investigate the reward and evaluation curves when training non-distilled models from scratch (i.e., Qwen2.5-Math-7B, Qwen2.5-32B, and Qwen3-30B-A3B). As shown in Figure~\ref{fig:convergence}, we compare \system with \reviewerB{VeRL, VeRL-Pipeline, and VeRL-Async}. In addition, we simulate the case of \reviewerB{completely removing staleness control} in both \system and VeRL-Async by disabling \system's global consistency protocol and removing the admission control that limits the amount of in-flight data in VeRL-Async, respectively.\footnote{\reviewerC{Without staleness control, all instances remain fully loaded continuously, causing \system's \textit{synchronization strategy} to never trigger model updates. Therefore, we} \reviewerC{instead adopt the vanilla model update strategy used by the other systems, where each instance updates to the latest model immediately after training completes.}}}

\reviewerC{The results reveal that:
(1) When staleness control is absent, or when the \textit{staleness bound} $\eta$ becomes excessively large (e.g., $\eta=10$), training collapses entirely. \reviewerB{This further demonstrates the importance of our global consistency protocol in guaranteeing convergence.}
(2) As long as the \textit{staleness bound} is properly configured, e.g., \system and VeRL-Async with $\eta = 3$, and VeRL-Pipeline (one-step staleness, $\eta = 1$), all systems exhibit convergence behavior comparable to fully synchronous training without staleness (i.e., VeRL, $\eta = 0$). This also suggests that \reviewerB{data skewness mitigation} (i.e., all rollout coordination introduced by \system over VeRL and other VeRL variants) \reviewerB{does not inherently affect convergence}. Instead, the \textit{staleness bound} is the primary determining factor.
(3) Under the same $\eta$, \system achieves convergence comparable to the baselines. For example, Figure~\ref{fig:convergence}(b) shows that both \system and VeRL-Async with $\eta = 3$ reach similar AIME24 scores after approximately 150 steps. However, \system delivers higher throughput than VeRL-Async, making it more efficient.}

\begin{figure}[!t]
    \centering
    \includegraphics[width=\linewidth]{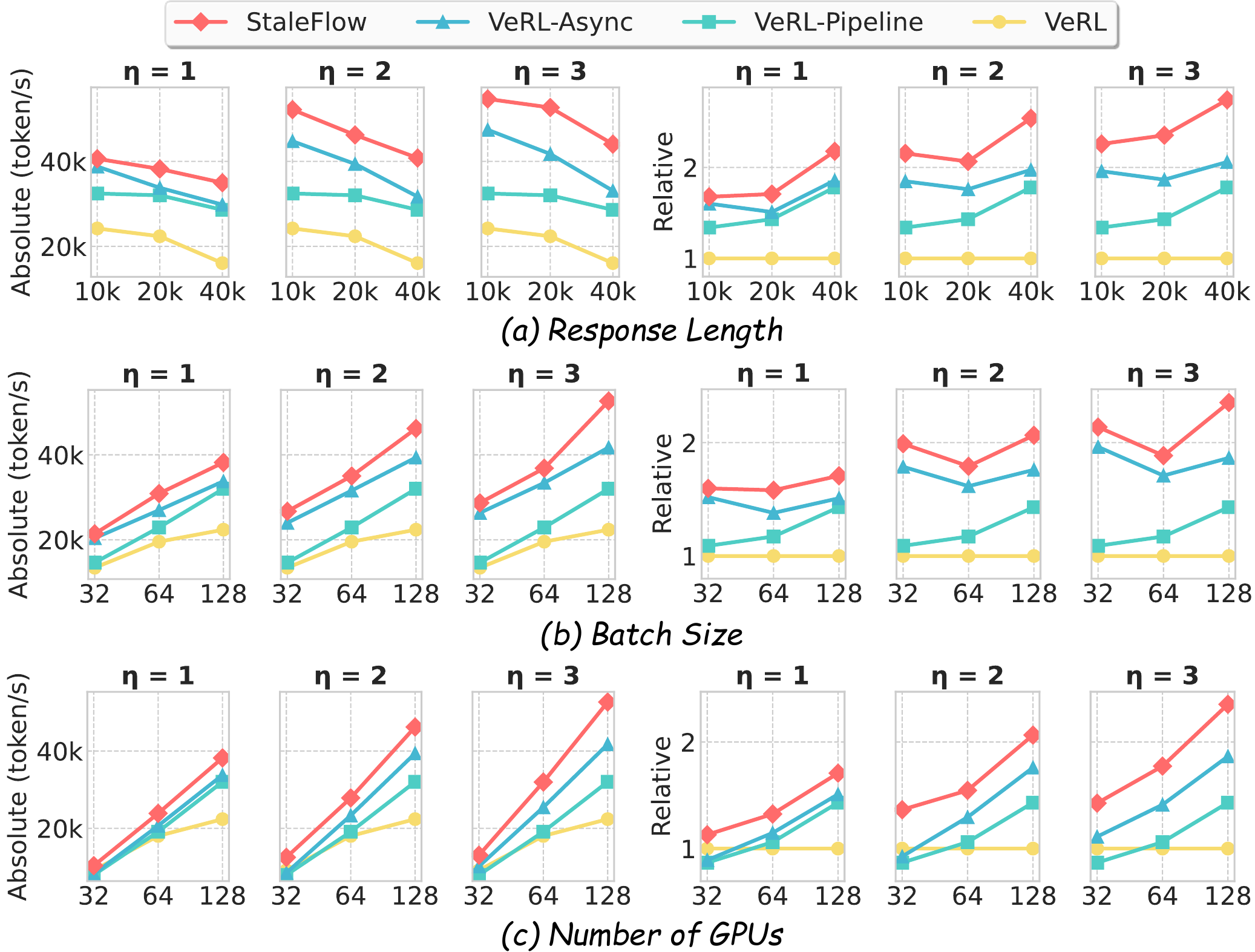}
    \myvspace{-18pt}
    \caption{\small{Absolute and relative throughput (normalized to VeRL) under different scaling factors.}}
    \label{fig:scalability}
    \myvspace{-5pt}
\end{figure}

\subsection{Scalability}
\label{subsec:scalability}

Next, we evaluate the scalability of \system compared to representative systems of different types. Among systems with strict staleness control, VeRL-Async delivers the best performance and is therefore selected. We use the Qwen3-30B-A3B model with a response length limit of 20K, batch size of 128, and 128 GPUs, consistent with the configuration in Figure~\ref{fig:e2e}(d). As shown in Figure~\ref{fig:scalability}, we vary one factor at a time  while holding the others constant.

\mysubsubsection{Response length} Longer response lengths exacerbate long-tail effects and increase data skewness, which generally reduces system throughput. Despite this, \system maintains the highest absolute throughput across all settings. More importantly, its throughput relative to VeRL scales better than that of other systems, underscoring its stronger ability to mitigate long-tail data skewness.

\mysubsubsection{Batch size} Increasing batch sizes improves GPU utilization and thereby boosts throughput. Compared to baselines, \system demonstrates superior scalability in both absolute throughput and relative gains. This indicates that StaleFlow translates larger batches into throughput improvement more efficiently, making better use of available trajectories per post-training step.

\mysubsubsection{Number of GPUs} Scaling the number of GPUs provides greater compute capacity, leading to higher throughput. \system scales at least comparably to other systems, and its advantage becomes more pronounced under larger \textit{staleness bounds}.

\subsection{Performance Interpretation}
\label{subsec:interpretation}

We further conduct a detailed analysis to identify the sources of throughput gain and provide a detailed breakdown of \system. The analysis uses a representative configuration: Qwen3-30B-A3B on 128 GPUs, with a batch size of 128, group size of 16, response length of 40K, and a \textit{staleness bound} of 3.

\mysubsubsection{Ablation study}
Corresponding to the three throughput-oriented rollout coordination strategies in \system, we implement three vanilla counterparts for comparison: (1) \textit{Vanilla routing.} Trajectories are assigned purely by load balancing their counts. Each trajectory in TS is routed to the instance with the fewest trajectories.
(2) \textit{Vanilla synchronization.} Model updates are performed greedily. An instance synchronizes the model immediately once a newer version is observed at the PS.
(3) \textit{Vanilla migration.} No proactive migration is applied. Trajectories migrate only passively upon synchronization interruptions, after being returned to the TS and re-routed.

\begin{figure}[!t]
    \centering
    \includegraphics[width=\linewidth]{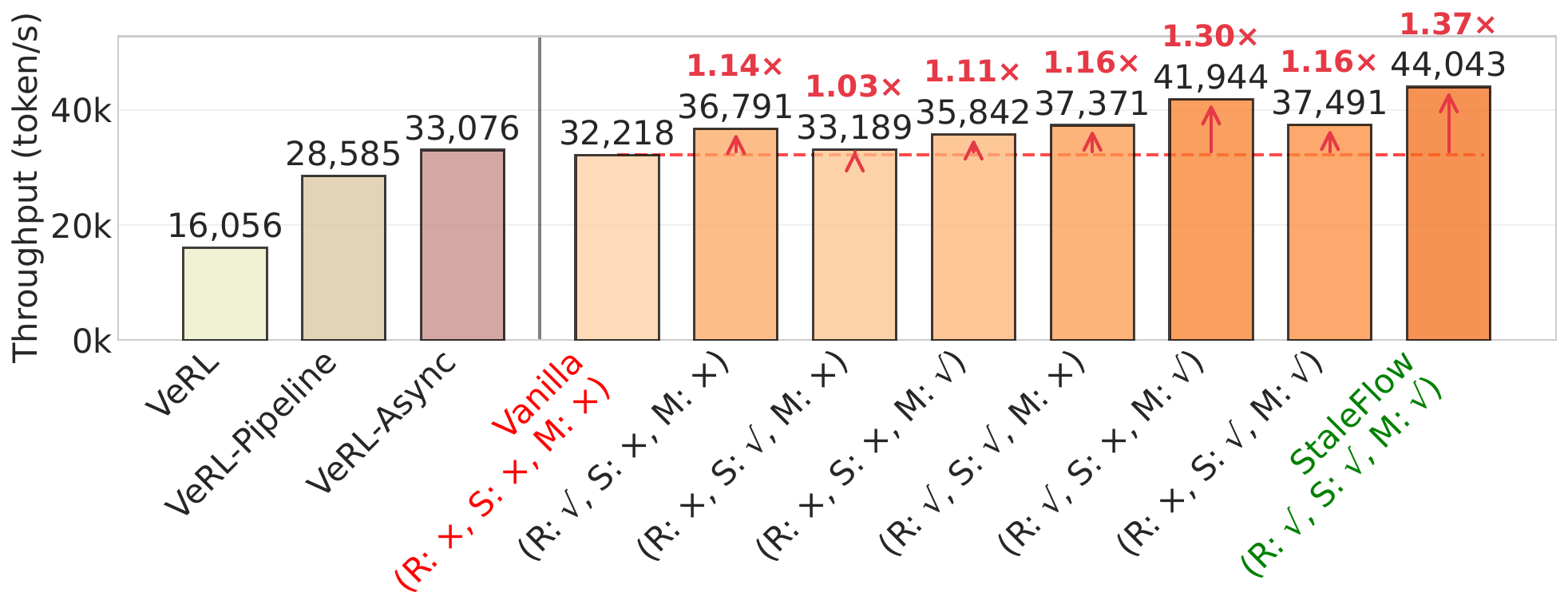}
    \myvspace{-23pt}
    \caption{\small{Ablation study of rollout coordination strategies. ``R'', ``S'', and ``M'' denote \textit{routing, synchronization, and migration strategies}, respectively. ``\checkmark'' indicates \system’s throughput-oriented strategy; ``$\times$'' indicates the vanilla counterpart.}}
    \label{fig:ablation}
    \myvspace{-5pt}
\end{figure}

Figure~\ref{fig:ablation} shows the results under different combinations of throughput-oriented and vanilla strategies. First, when all vanilla strategies are used, performance closely matches that of VeRL-Async, confirming that the gains of \system do not stem from engineering artifacts. Second, progressively replacing vanilla strategies with those of \system consistently improves throughput, with the highest gain achieved when all three strategies are enabled. We interpret these results as follows.
(1) \textit{Vanilla routing} balances trajectory counts but fails to balance or optimize per-instance throughput.
(2) \textit{Vanilla synchronization} triggers more frequent model updates, which interrupt more ongoing trajectories and incur extra KV Cache recomputation overhead. In contrast, \system synchronizes per instance based on actual load, significantly reducing this overhead.
(3) \textit{Vanilla migration} lacks proactive redistribution, making it hard to rebalance the workload once skewness occurs.

\begin{figure}[!t]
    \centering
    \includegraphics[width=\linewidth]{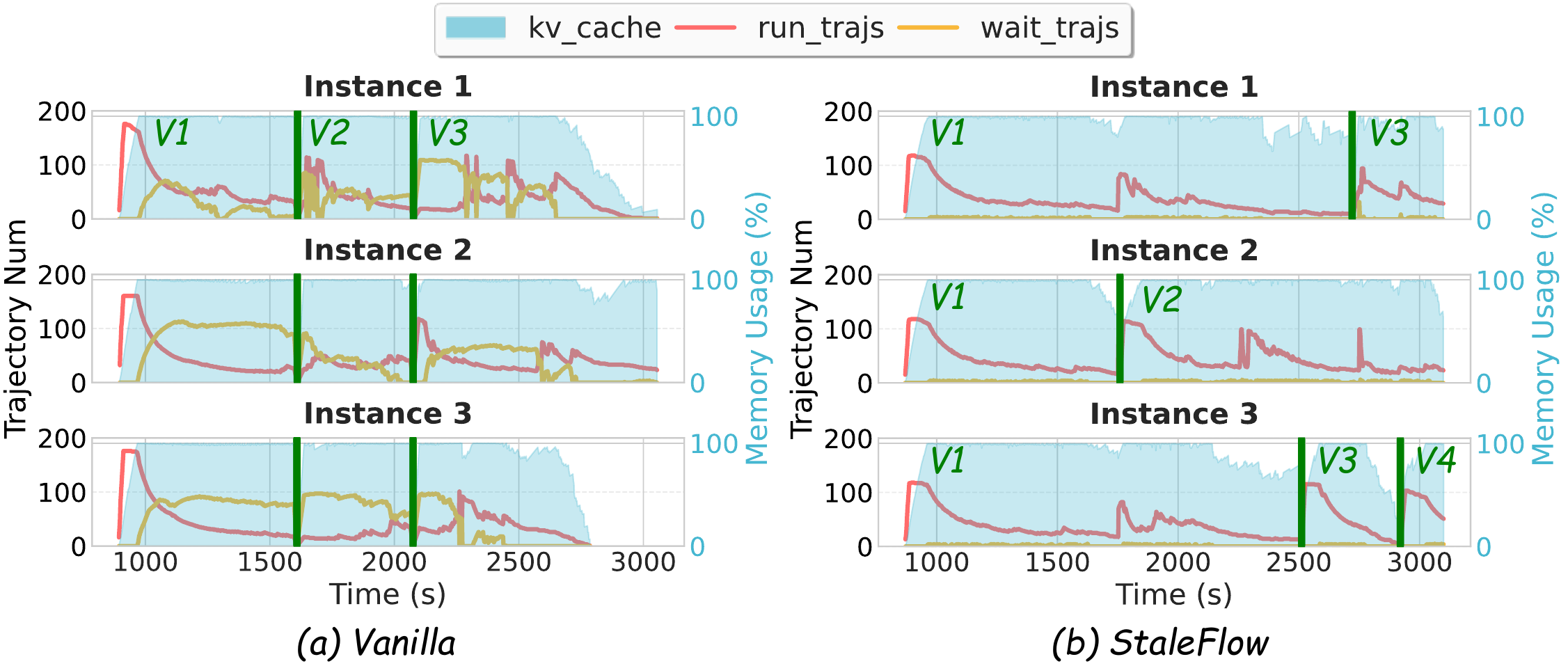}
    \myvspace{-20pt}
    \caption{\small{Per-Instance rollout load over time. The green lines indicate model synchronizations and tags indicate model versions.}}
    \label{fig:case}
    \myvspace{-5pt}
\end{figure}

\mysubsubsection{Case study}
To better illustrate the effect of each strategy, we present  the per-instance rollout load over 3000 seconds when using all \system strategies versus all vanilla counterparts. As shown in Figure~\ref{fig:case}, we observe: (1) \textit{Vanilla routing} distributes all assignable trajectories evenly across instances, while \system routes based on the marginal throughput gain threshold $\mu$. For example, in Figure~\ref{fig:case}(b), each instance initially receives only about 100 trajectories, with the rest kept at the TS. This leaves more flexibility for subsequent adjustments when load imbalance emerges. (2) \textit{Vanilla synchronization} performs greedy global updates even while many trajectories are still running. In Figure~\ref{fig:case}(a), instance~2 updates from $V_2$ to $V_3$ during heavy execution, incurring extra KV Cache recomputation overhead. \system, however, synchronizes per instance, following each instance's own pace. (3) \textit{Vanilla migration} does not proactively redistribute trajectories, leading to growing waiting queues and load imbalance, whereas \system continuously migrates trajectories via the TS, preventing this imbalance.

\begin{figure}[!t]
    \centering
    \includegraphics[width=\linewidth]{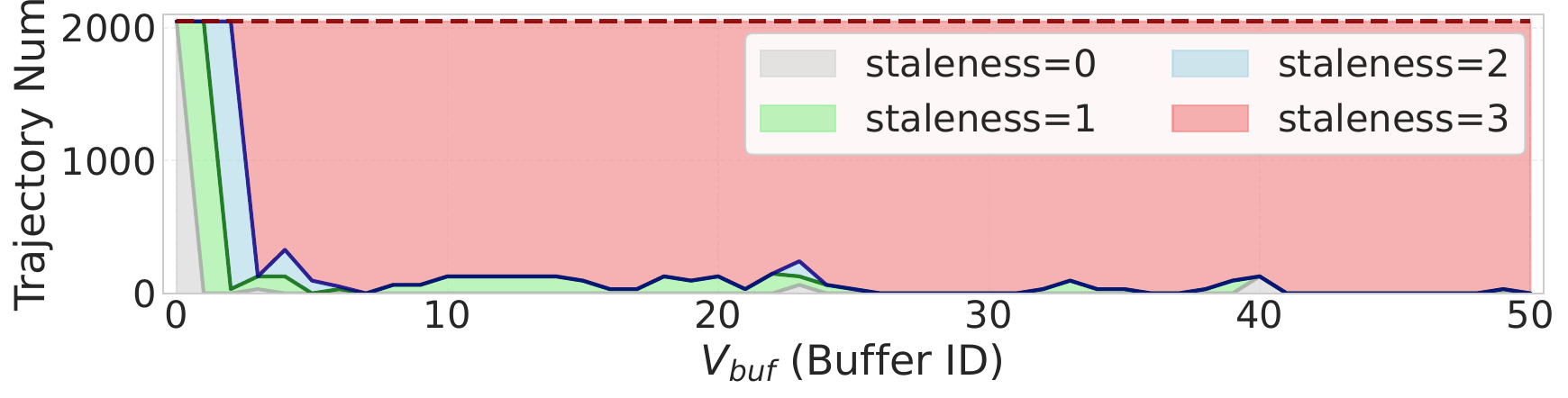}
    \myvspace{-23pt}
    \caption{\small{Trajectory staleness distribution across different staleness buffers. The \textit{staleness bound} $\eta$ is set to 3.}}
    \label{fig:buffer}
    \myvspace{-5pt}
\end{figure}

\mysubsubsection{Staleness distribution} We next investigate the staleness distribution of different trajectories in each staleness buffer (i.e, $V_{buf} - V_{traj}$). As shown in Figure~\ref{fig:buffer}, we observe that:
(1) No trajectory exceeds a staleness of 3, meaning all trajectories satisfy the constraint $V_{traj} + 3 \ge V_{buf}$. This reflects the effectiveness of our global consistency protocol in achieving strict staleness control while supporting flexible rollout coordination.
(2) As training proceeds, most buffers contain only trajectories with staleness 3. This indicates that \system fully exploits the maximum bound $\eta$ to squeeze out system throughput by tolerating higher staleness where possible.

\begin{figure}[!t]
    \centering
    \includegraphics[width=\linewidth]{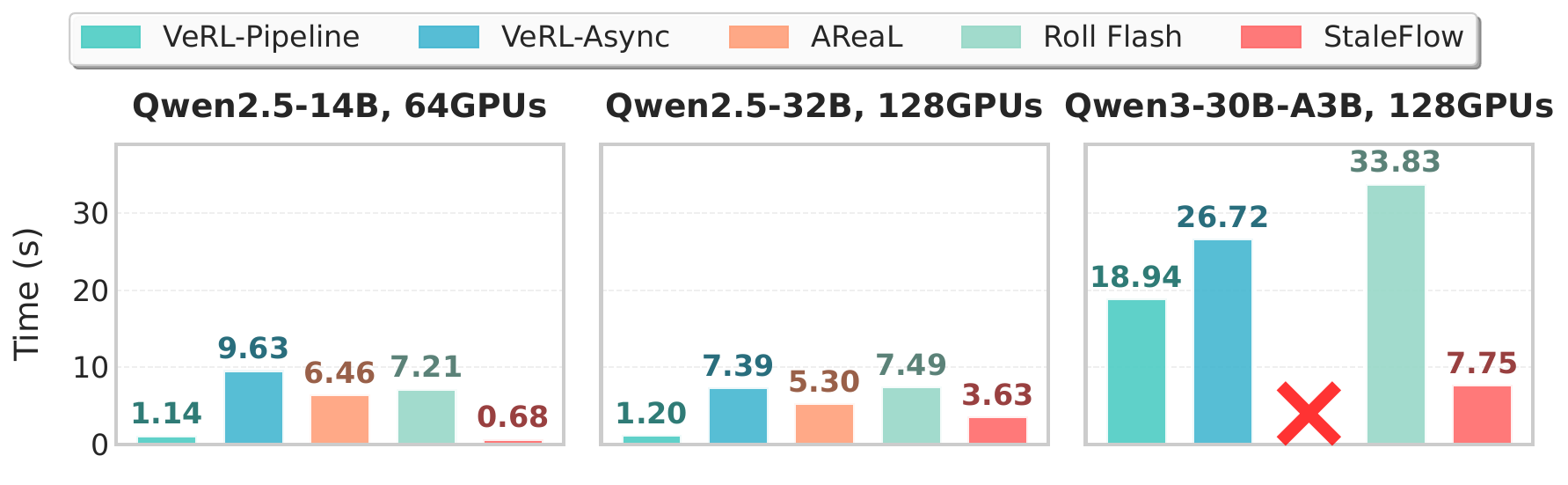}
    \myvspace{-20pt}
    \caption{\small{Overhead of model synchronization.}}
    \label{fig:sync}
    \myvspace{-5pt}
\end{figure}

\mysubsubsection{Time breakdown} 
As presented in Table~\ref{tab:time_breakdown}, we report the average per-step rollout time, which also constitutes the total RL post-training step time (since the reward and training phases are overlapped by the rollout phase).  It can be observed that: \reviewerB{(1)} Per-token decoding during trajectory generation accounts for the majority of the time (\reviewerB{89.2\%}). Meanwhile, KV Cache prefill or re-prefill after trajectory routing or migration takes 7.9\%. \reviewerB{(2) The overhead of our global consistency protocol (including \texttt{Reserve} and \texttt{Occupy}) remains below 1\%. (3) The overhead introduced by our rollout coordination commands is also below 3\%.} Specifically, the \texttt{Route} and \texttt{Interrupt} commands for trajectories only involve communicating lightweight tokens, thus consuming merely 0.5\%, whereas \texttt{Pull} requires transferring larger model parameters and hence takes longer, but still only 1.7\%. Furthermore, we compare the model synchronization time under different configurations with other baselines, as shown in Figure~\ref{fig:sync}. \system achieves the minimal latency in most settings. This demonstrates that the additionally introduced TS and PS, along with the various rollout commands designed atop them, introduce negligible overhead.\footnote{In addition, the overhead of Algorithm~\ref{alg:overall} is typically under 0.1s, which continuously runs in the background process of the rollout coordinator.}

\begin{table}[!t]      
  \centering
  \caption{\small{Rollout execution time breakdown (per-step).}}                    
  \myvspace{-10pt}
  \small
  \begin{tabular}{ccccccc}
  \toprule
  \multicolumn{3}{c}{{Rollout Coordination}} & \multicolumn{2}{c}{\reviewerB{Consistency Protocol}} &
  \multicolumn{2}{c}{{Generation}} \\
  \cmidrule(lr){1-3} \cmidrule(lr){4-5} \cmidrule(lr){6-7}
  \texttt{Pull} & \texttt{Route} & \texttt{Interrupt} & \texttt{\reviewerB{Reserve}} & \texttt{\reviewerB{Occupy}} & \text{Prefill} &
  \text{Decode} \\
  \midrule
  7.8s & 2.0s & 0.6s & \reviewerB{0.8s} & \reviewerB{2.2s} & 36.4s & 413.3s \\
  (1.7\%) & (0.4\%) & (0.1\%) & \reviewerB{(0.2\%)} & \reviewerB{(0.5\%)} & (7.9\%) & (89.2\%) \\
  \bottomrule
  \end{tabular}
  \label{tab:time_breakdown}
  \myvspace{-5pt}
\end{table}

%% file: sections/conclusion.tex
\section{Related Work}
\label{sec:related_work}

\mysubsubsection{RL post-training systems}
A number of RL post-training systems have emerged recently. Early systems largely adopted synchronous architectures with shared resources~\cite{verl, rlhfuse, flexrlhf, real, puzzle, roll, openrlhf, foremoe}, while more recent efforts have shifted toward asynchronous and disaggregated designs~\cite{stream_rl, verl_pipeline, llama_rl, areal, roll_flash, rhy_rl, laminar, async_rlhf, asyncflow}. These works propose various rollout coordination techniques to mitigate data skewness (see \S\ref{subsec:data_skewness}). However, most of them are ad hoc, and none of the existing systems provides a comprehensive staleness control protocol. As a result, they cannot perform holistic global optimization to mitigate data staleness under a bounded staleness. 

In addition, several works focus on improving the efficiency of rollout itself, for example, by reducing generation latency via speculative decoding~\cite{seer, specrl, rhy_rl, respec, beat_long_tail} or low-precision quantization~\cite{qerl, roll_flash}. These efficient RL optimizations are orthogonal to our work and can be seamlessly incorporated into \system.

\mysubsubsection{Optimizing training and generation}
Significant progress has been made in training through hybrid parallelisms, including data parallelism~\cite{pytorch_fsdp, zero}, tensor parallelism~\cite{megatron_1, megatron_2}, pipeline parallelism~\cite{gpipe, pipedream, dynapipe, pipeline-overview}, sequence parallelism~\cite{sp, cp}, and expert parallelism~\cite{gshard}. Systems such as Megatron~\cite{megatron_1, megatron_2} and DeepSpeed~\cite{zero, sp} have integrated multiple parallelisms, while many works~\cite{galva, galva-pipe, hetu, malleus, elastor} further enable automatic parallelization tuning. For model generation, numerous optimizations have been proposed, including speculative decoding~\cite{spec}, chunked prefill~\cite{chunk_prefill}, prefill-decode disaggregation~\cite{distserve}, prefix caching~\cite{hotprefix}, KV Cache quantization~\cite{pqcache}, and other management strategies~\cite{aptserve, dali}. Many of these techniques have been integrated into systems such as vLLM~\cite{vllm} and SGLang~\cite{sglang}.

All these training and generation optimizations are applicable to \system. By adopting a fully disaggregated architecture, \system cleanly decouples rollout (generation) and training, allowing each RL phase to be optimized independently.

\mysubsubsection{Optimizing data skewness}
Data skewness caused by variable-length sequences is a well-known problem for large models. Existing systems have tackled it during model pre-training~\cite{hotspa, hydraulis, flexsp, llm-dataset}, fine-tuning~\cite{long_align, lobra}, and serving~\cite{long_serve}, typically by dynamically adapting parallelism strategies to accommodate varying lengths. In contrast, \system targets data skewness specifically in RL post-training. Compared to other scenarios, RL introduces an additional challenge: data staleness, which does not arise in conventional training or serving pipelines. It is tightly coupled with data skewness and must be optimized jointly. This makes our target setting fundamentally more complex, as \system must prioritize convergence while maximizing system performance. Consequently, \system incorporates staleness constraints into its data management design.

\section{Conclusion and Future Work}
This work presents \system, a fully disaggregated RL post-training system that jointly addresses data staleness and data skewness. \system introduces a global consistency protocol that enables fine-grained staleness control, together with architectural and algorithmic innovations that support staleness-aware, throughput-oriented rollout coordination strategies for mitigating skewness. While preserving RL convergence, \system achieves up to 1.42-2.68$\times$ (\reviewerC{1.18-1.91}$\times$ on average) higher throughput than state-of-the-art RL post-training systems. As future work, we plan to extend \system to support a broader range of RL tasks (e.g., coding, agent-based~\cite{agent-rise, llm-agents, llm-tool}) and a wider spectrum of models (e.g., multi-modal~\cite{multi-modal}). Since \system is agnostic to the types of tasks and models, it is broadly applicable.

%% file: sections/appendix.tex
\section{Implementation}
\label{appendix:impl}

\system is implemented in approximately 22k lines of code (LoC) in Python. (1) The staleness manager, which enforces the global consistency protocol, accounts for around 2k LoC. (2) On the training side, \system supports the Megatron~\cite{megatron_1, megatron_2} and FSDP2~\cite{pytorch_fsdp} backends (Megatron is used in our evaluation) and consists of nearly 2k LoC. (3) The rollout side (i.e., the rollout service) consists of roughly 10k LoC, with approximately 3k LoC dedicated to the core logic of rollout generation using vLLM~\cite{vllm}, 2k LoC for the rollout coordinator, 1k LoC for the trajectory server (TS), and 4k LoC for the parameter server (PS).

For efficient communication of parameters on the PS, we leverage the NVIDIA Inference Xfer Library (NIXL)~\cite{nixl}. NIXL builds on Unified Communication X (UCX)~\cite{ucx}, which abstracts point-to-point (P2P) communication across various hardware (GPU/CPU/Storage), hiding the memory and communication details specific to each hardware. This abstraction enables the PS to be deployed on CPUs while ensuring efficient communication with GPUs used in both training and rollout. In the following, we provide a detailed walkthrough of the PS deployment (Appendix~\ref{appendix:deploy}), its communication optimizations (Appendix~\ref{appendix:comm}) and its scalability analysis (Appendix~\ref{appendix:comm_scale}).

\subsection{Parameter Server Deployment}
\label{appendix:deploy}

\begin{figure}[!t]
\centering
\includegraphics[width=\linewidth]{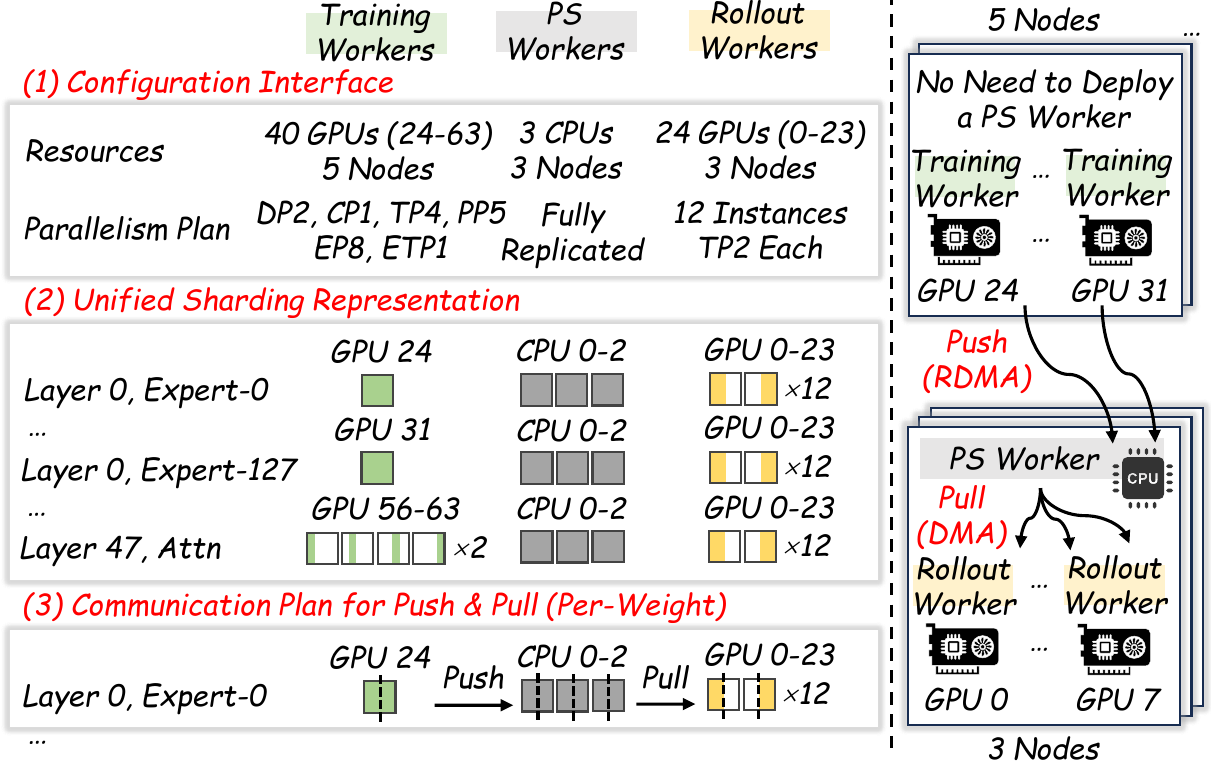}
\caption{\small{Deployment of the Qwen3-30B-A3B model across 64 H20 GPUs. (Left) The PS adopts a worker-based architecture consistent with both the training and rollout phases, supporting flexible resource placement and parallelism. The labels ``DP, CP, TP, PP, EP, ETP'' refer to different parallelisms—data, context, tensor, pipeline, expert, and expert-specific tensor parallelism—that collectively determine how model parameters are distributed and assigned to resources. (Right) The illustration of the deployment strategy used for \texttt{Pull} and \texttt{Push}.}}
\label{fig:ps_appendix}
\end{figure}

\begin{figure}[!t]
\centering
\includegraphics[width=\linewidth]{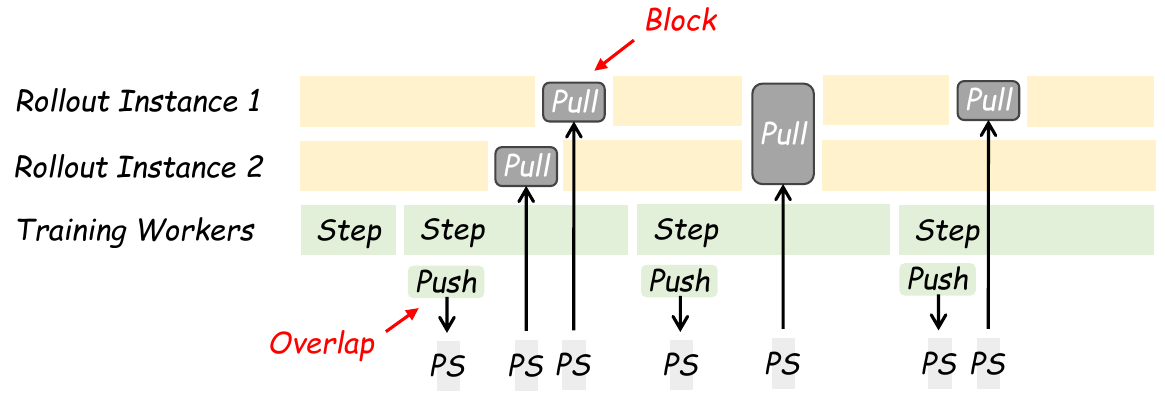}
\caption{\small{\texttt{Push} is triggered automatically by training workers and can overlap with the next training step, whereas \texttt{Pull} is issued by the rollout coordinator and must block the generation of target instances.}}
\label{fig:overlap_push}
\end{figure}

\begin{figure}[!t]
\centering
\includegraphics[width=\linewidth]{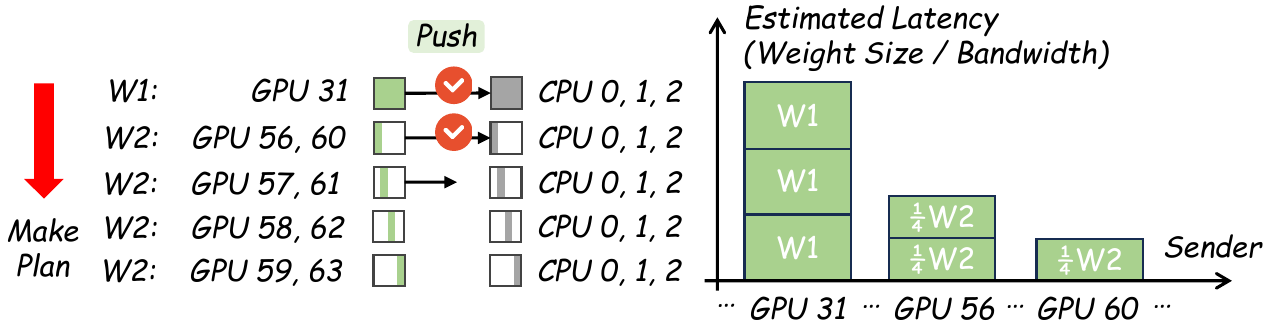}
\caption{\small{Load-balancing communication planning for \texttt{Push}. The estimated latency is computed by dividing the data size of the current slice (in bytes) by the bandwidth between the sender and receiver (in bytes/s), plus a constant latency term (in s). Both the bandwidth and the constant latency term can be profiled in advance.}}
\label{fig:comm}
\end{figure}

The PS operates as a distributed data plane that utilizes only CPU resources, serving as middleware between the training and rollout phases. As illustrated in Figure~\ref{fig:ps_appendix} (left), the deployment follows a three-step process: (1) The PS is instantiated as a collection of workers, with resource allocation and parallelism plans specified through a unified interface, which is also used to configure training and rollout workers. This configuration defines how model parameters are partitioned among resources according to the parallelism plan. (2) A converter standardizes the various parallelism representations used by the training, PS, and rollout phases into \system's unified sharding representation, ensuring consistency in parameter names and slicing granularity. (3) Synchronization is achieved through \texttt{Push} operations from training workers to PS workers and \texttt{Pull} commands from PS workers to rollout workers. Both of them are governed by load-balancing communication plans (detailed in Appendix~\ref{appendix:comm}) that optimize the sending and receiving of parameter slices, minimizing overall communication latency.

As shown in Figure~\ref{fig:overlap_push}, \texttt{Pull} blocks the rollout generation. However, \texttt{Push} can overlap with the subsequent training step, since it only needs to complete before the optimizer applies the parameter updates. Consequently, the non-overlapped cost of \texttt{Pull} becomes a more critical performance concern. Motivated by this, as shown in Figure~\ref{fig:ps_appendix} (right), we deploy PS workers on the same nodes as rollout workers and use the fully replicated strategy, so that \texttt{Pull} only involves PCIe DMA copy, which is significantly faster, while \texttt{Push} leverages RDMA across nodes.

\subsection{Load-Balancing Communication Plan}
\label{appendix:comm}

The communication plan governs how model parameter slices are communicated across resources during \texttt{Push} and \texttt{Pull}. As shown in Figure~\ref{fig:comm}, we describe the process using \texttt{Push} as an example. 

To formulate the \texttt{Push} communication plan, we begin by aligning the smallest granularity of parameter slicing. With this common slicing granularity established, the next step is to determine optimal P2P pairs between senders and receivers for each parameter slice. Here, a single slice may originate from multiple potential senders but must reach one or more specific receivers. To balance the communication load across senders, we track for each sender an accumulated estimate of the communication latency its assigned slices would incur. When assigning a sender for a receiver’s required slice, our planner selects the candidate sender with the smallest current accumulated latency, thereby greedily distributing traffic to minimize bottlenecks. Once obtained, this set of sender-receiver mappings is kept static and reused for every subsequent \texttt{Push}.

\subsection{Scalability}
\label{appendix:comm_scale}

\begin{figure}[!t]
\centering
\includegraphics[width=0.8\linewidth]{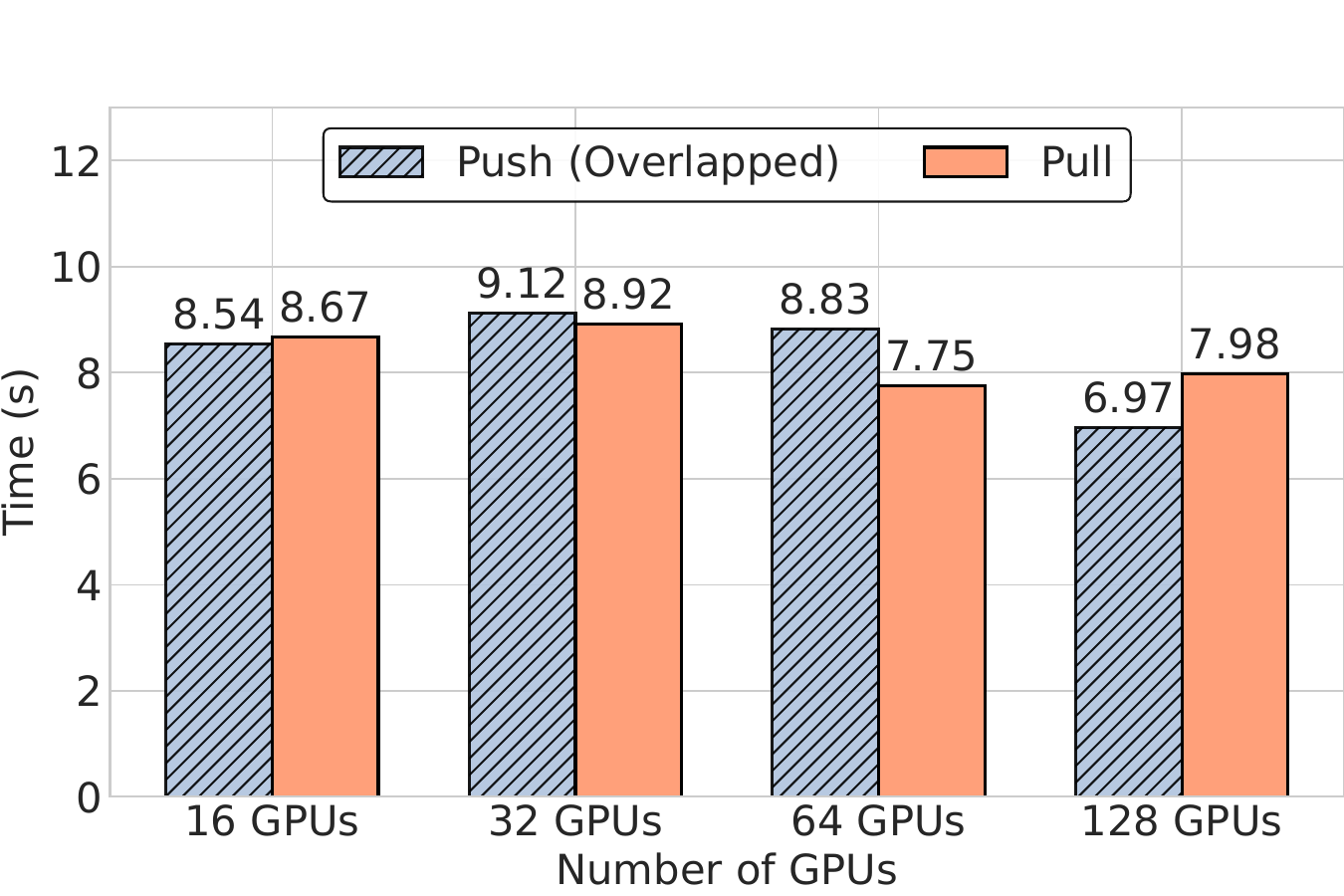}
\caption{\small{Communication overhead across varying cluster scales.}}
\label{fig:sync_scale}
\end{figure}

We evaluate the scalability of model parameter synchronization under our PS architecture. Using the Qwen3-30-A3B model, we measure the overhead of the \texttt{Push} (which is overlapped with training) and \texttt{Pull} across 16 to 128 GPUs. As shown in Figure~\ref{fig:sync_scale}, increasing the cluster size does not lead to higher communication overhead. This is because, as the cluster expands, the number of rollout workers, training workers, and PS workers increases proportionally, with additional model replicas. The optimal communication plans for small and large-scale setups exhibit similar patterns: replicas communicate independently and concurrently, balancing the load without interference. Therefore, proportionally scaling the number of model replicas does not affect communication overhead, demonstrating that the PS design exhibits good scalability.

\begin{figure*}[!t]
\centering
\includegraphics[width=0.8\linewidth]{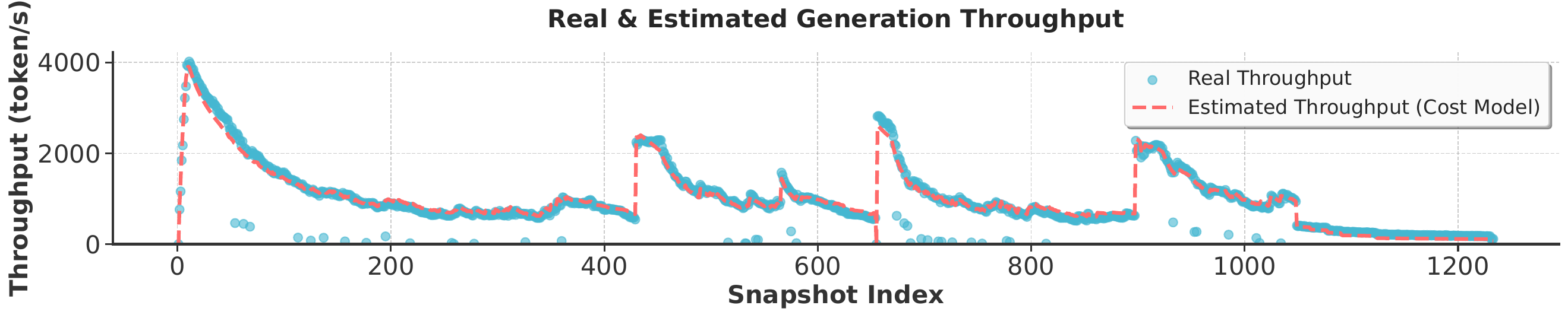}
\caption{\small{Comparison of actual throughput and estimated throughput from our cost model. We present the results for a single instance over the first two post-training steps.}}
\label{fig:acc}
\end{figure*}

\begin{figure*}[!t]
\centering
\includegraphics[width=0.8\linewidth]{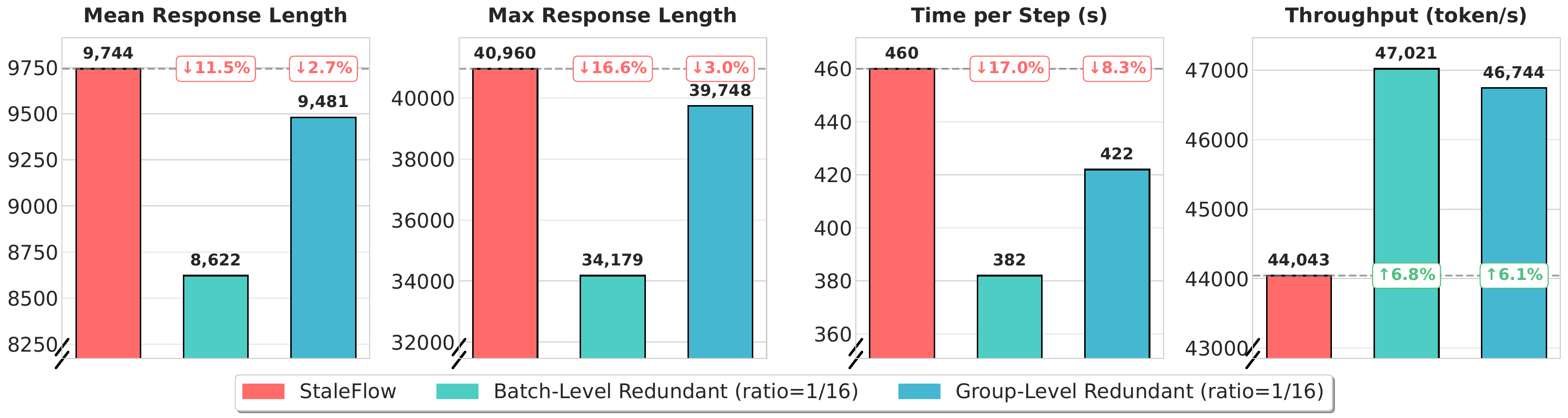}
\caption{\small{Effect of redundant rollout. We compare \system with and without redundant rollout at the batch and group levels. Green bars represent performance improvements, while red bars indicate performance degradation.}}
\label{fig:redundant}
\end{figure*}

\section{Cost Model}
\label{appendix:cost_model}

In this section, we present the detailed derivation of our cost model $\mathcal{T}_i(S)$, which represents the estimated generation throughput of instance $i$ given the snapshot $S$ (Appendix~\ref{appendix:derivation}). We then compare these estimates with the actual measured throughput to assess the accuracy of our cost model (Appendix~\ref{appendix:acc}).

\subsection{Derivation}
\label{appendix:derivation}

Since decoding occupies most of the time during rollout (see Table~\ref{tab:time_breakdown}), we simplify the throughput estimation by treating the pure decoding throughput as the overall generation throughput. It can be obtained by dividing the number of running trajectories by the per-decoding-step latency.

Following prior work~\cite{distserve}, we model the per-decoding-step latency as the sum of the latencies of its primary computational components. The dominant components are the attention operation and the feed-forward network (FFN), which primarily consist of multiple matrix multiplications.

\begin{table}[!t]
\small
\centering
\caption{\small{Coefficients of the cost model for Qwen3-30B-A3B. These coefficients are derived via offline profiling and utilized for estimating the system's throughput and marginal throughput gain.}}
\begin{tabular}{cc}
\toprule
Coefficients & Values \\
\midrule
$k_1$ & $7.28 \times 10^{-8}$ \\
$k_2$ & $1.72 \times 10^{-3}$ \\
$k_3$ & $1.25 \times 10^{-4}$ \\
$k_4$ & $1.07 \times 10^{-2}$ \\
\bottomrule
\end{tabular}
\label{tab:coefficient}
\end{table}

\mysubsubsection{Latency of a single operator}
In GPU execution, the latency of an operator is determined by the slower of its memory access latency and its computation latency:
\begin{equation}
\small
\label{eq:op_latency}
\mathcal{L} = \max(\text{memory\_latency}, \text{computation\_latency}).
\end{equation}

\mysubsubsection{Attention latency}
The attention operation (e.g., as implemented in FlashInfer~\cite{flash_infer}) is typically memory-bandwidth bound. Its latency is dominated by the time required to read the KV Cache from GPU memory. We model it as linearly proportional to the size of the KV Cache for the instance:
\begin{equation}
\small
\label{eq:attn_latency}
\mathcal{L}_{{attn}} = k_1 \times \text{kv\_cache}.
\end{equation}
Here, $k_1$ is a constant coefficient representing the inverse effective memory bandwidth for accessing the KV Cache. $\text{kv\_cache}$ denotes the total size (e.g., in bytes) of the cached keys and values for all running trajectories being processed by the instance.

\mysubsubsection{Matrix multiplication latency}
The FFN and projection layers mainly consist of matrix multiplications. Consider a general matrix multiplication of the form $\mathbf{Y} = \mathbf{X}\mathbf{W}$, where $\mathbf{X} \in \mathbb{R}^{n \times h}$ is the input ($n$ is the number of running trajectories, $h$ is the hidden dimension) and $\mathbf{W} \in \mathbb{R}^{h \times h'}$ is the parameter matrix.

\begin{itemize}[noitemsep, topsep=0pt, parsep=0pt, partopsep=0pt, leftmargin=*]
    \item \textbf{Memory latency:} The dominant memory cost for this operator is reading the parameter matrix $\mathbf{W}$ from GPU High-Bandwidth Memory (HBM), as inputs are typically already cached. This cost is proportional to the size of $\mathbf{W}$:
    \begin{equation}
    \small
    \text{memory\_latency}_{{matmul}} = A \times (h \times h').
    \end{equation}
    $A$ is a constant related to the inverse memory bandwidth for parameter access.

    \item \textbf{Computation latency:} The number of floating-point operations (FLOPs) is $2 \times n \times h \times h'$. Therefore, the computation latency is proportional to this:
    \begin{equation}
    \small
    \text{computation\_latency}_{{matmul}} = B \times (2 \times n \times h \times h').
    \end{equation}
    $B$ is a constant related to the inverse computational throughput (FLOP/s) of the GPU.
\end{itemize}
Applying Eq.~\ref{eq:op_latency}, the latency of this matrix multiplication is:
\begin{equation}
\small
\label{eq:matmul_latency_raw}
\mathcal{L}_{{matmul}} = \max\left( A \times (h \times h'), \; B \times (2 \times n \times h \times h') \right).
\end{equation}
This can be simplified by defining $k_2 = A \times (h \times h')$ and $k_3 = B \times (2 \times h \times h')$, yielding:
\begin{equation}
\small
\label{eq:matmul_latency}
\mathcal{L}_{{matmul}} = \max( k_2,\; k_3 \times n ).
\end{equation}

The critical point where the bottleneck shifts from memory to computation occurs when $k_3 \times n > k_2$, i.e., $n > k_2 / k_3$. The ratio $k_2/k_3$ represents the \textit{arithmetic intensity} threshold (in FLOPs/byte) required for the operation to become compute-bound. For example, with bf16 precision on an H20 GPU, this threshold is empirically found to be approximately 37. Thus, for $n > 37$, $\mathcal{L}_{{matmul}} \approx k_3 \times n$.

\mysubsubsection{Total per-decoding-step latency}
Combining the latency of attention and the dominant matrix multiplication, along with a small constant overhead $k_4$ for other operations (e.g., activation functions, layer normalization) and other latencies like kernel launch, the total per-decoding-step latency for an instance $i$ is:
\begin{equation}
\small
\label{eq:total_latency}
\begin{aligned}
\mathcal{L}_i &= \mathcal{L}_{{attn}} + \mathcal{L}_{{matmul}} + k_4 \\
    &= k_1 \times \text{kv\_cache}_i + \max\left( k_2,\; k_3 \times n_i \right) + k_4.
\end{aligned}
\end{equation}
Here, $n_i$ represents the number of running trajectories in instance $i$.

\mysubsubsection{Instance generation throughput}
Therefore, the generation throughput (tokens per second) for instance $i$ is the number of running trajectories divided by the per-decoding-step latency, as each step generates one token per trajectory in a batched decode:
\begin{equation}
\small
\label{eq:throughput_raw}
\mathcal{T}_i = \frac{n_i}{\mathcal{L}_i} = \frac{n_i}{k_1 \times \text{kv\_cache}_i + \max\left( k_2,\; k_3 \times n_i \right) + k_4}.
\end{equation}

To align with the paper's notation, we use the snapshot $S[i]$, where its field $S[i].\text{run\_trajs}$ gives the set of running trajectories for instance $i$, and thus $n_i = |S[i].\text{run\_trajs}|$. The KV Cache memory consumption for the instance is similarly referenced as $S[i].\text{kv\_cache}$.

Thus, we obtain the final throughput estimation presented in the paper:
\begin{equation}
\small
\label{eq:throughput_final}
\boxed{
\mathcal{T}_{i}(S)= \frac{|S[i].\text{run\_trajs}|}{k_1 \times S[i].\text{kv\_cache} + \max(k_2,\; k_3 \times |S[i].\text{run\_trajs}|) + k_4}.}
\end{equation}
The coefficients $k_1$ through $k_4$ are determined via offline profiling of the rollout engine and linear regression.

\subsection{Estimation Accuracy}
\label{appendix:acc}

We assess the accuracy of our cost model, $\mathcal{T}_i(S)$, by comparing the estimated generation throughput with the actual throughput observed in our experiments. Figure~\ref{fig:acc} presents the results. The experimental setup aligns with \S\ref{subsec:interpretation}, using Qwen3-30B-A3B on 128 GPUs, a batch size of 128, a group size of 16, a response length of 40K, and a \textit{staleness bound} of 3. The constant coefficients used in the cost model are profiled and summarized in Table~\ref{tab:coefficient}. 

As shown in the figure, our cost model provides an accurate estimation of the actual throughput, with only a few outliers. The average estimation error is 10.52\%, demonstrating that the model can reliably guide the subsequent rollout coordination strategies to tune system throughput.

\section{Ablation Study on Redundant Rollout}
\label{appendix:redundant}

This section investigates the impact of redundant rollout on system performance. As illustrated in \S\ref{subsec:staleness_advance} and Figure~\ref{fig:staleness_advance}(b), redundant rollout involves generating more trajectories than needed for a given batch size or group size, then aborting the excess once the batch or group is full. This technique helps filter out excessively long or time-consuming trajectories, reducing data skewness and improving overall system throughput.

We use the same experimental setup as in \S\ref{subsec:interpretation}: Qwen3-30B-A3B on 128 GPUs, with a batch size of 128, group size of 16, response length of 40K, and a \textit{staleness bound} of 3. To quantify the redundancy, we define the \textit{redundant ratio}, which specifies the fraction of extra trajectories generated beyond the required number for training. For example, with a \textit{redundant ratio} of 1/16, for a batch size of 128 and a group size of 16, batch-level redundant rollout generates $128 + 128 / 16 = 136$ trajectory groups per step, while group-level redundant rollout generates $16 + 16 / 16 = 17$ trajectories per group. In both cases, the total number of trajectories per step increases to $128 \times 16 \times (1 + 1/16) = 2176$.

Figure~\ref{fig:redundant} illustrates the effects of batch-level and group-level redundant rollout with a \textit{redundant ratio} of 1/16. We observe a noticeable reduction in both the mean and maximum response lengths per step, with a more significant effect when using batch-level redundancy. This occurs because trajectories within a group tend to be either all long or all short; although the \textit{redundant ratio} is the same, batch‑level redundancy exerts a stronger impact as it may discard an entire group consisting of long sequences.

Due to the overall shortening of sequences, the number of tokens processed per step decreases, leading to a clear reduction in post‑training time per step. Besides, system throughput, calculated as the total number of tokens divided by the total time, improves modestly. Although both total tokens and total time decrease, redundant rollout helps eliminate excessively long trajectories, thereby reducing overall data skewness and thus improving throughput.

The experiments in \S\ref{sec:exp} do not employ this technique because, as we observed, it introduces a noticeable distribution shift in the generated length. To ensure a fair comparison with the baselines, we choose to discuss it separately here as an ablation study.

\section{Detailed Algorithms for Rollout Coordination Strategies}
\label{appendix:alg}

In this section, we present the detailed algorithms for all rollout coordination strategies introduced in \S\ref{subsec:strategy}: the \textit{routing strategy} (Algorithm~\ref{alg:routing}), the \textit{synchronization strategy} (Algorithm~\ref{alg:sync}), and the \textit{migration strategy} (Algorithm~\ref{alg:migration}).
These strategies rely on a helper function that determines whether a given trajectory $\tau$ can be assigned to a specific rollout instance $i$ without violating the staleness constraint. This function is implemented as shown in Algorithm~\ref{alg:check_routable}.
Collectively, these algorithms enable staleness-aware, throughput-oriented rollout coordination, effectively mitigating data skewness and enhancing overall system performance.

\begin{algorithm}[!h]
\caption{Check routable.}
\label{alg:check_routable}
\small
\KwIn{$S$: snapshot, $\tau$: trajectory to be routed, $i$: target instance}
\KwOut{Boolean indicating if  trajectory $\tau$ can be routed to instance $i$ without violating the staleness constraint}
    \uIf{$\tau.V_{traj}$ = \text{None}}{
        $\tau.V_{traj} \gets S[i].\text{inst\_version}$ \; 
        \If{$\text{staleness\_manager\_verify}(\tau.V_{traj})$}{
            \Return \text{True} \;
        }
    }
    \Else{
        \If{$S[i].\text{inst\_version} \ge \tau.V_{traj}$}{
            \Return \text{True} \;
        }
}
\Return \text{False} \;
\end{algorithm}

\newpage

\begin{algorithm}[!h]
\caption{Routing strategy.}
\label{alg:routing}
\small
\KwIn{$S$: snapshot of rollout instances, $\text{ts\_trajs}$: trajectories in TS}
\KwOut{List of (instance, trajectories) pairs for routing}
$\text{routing} \gets \emptyset$ \tcp*{Routing list}
$\text{MLQ} \gets \text{sort}(\text{ts\_trajs})$ \tcp*{Order by $V_{traj}$ ascending}
$\text{stop} \gets \text{False}$ \tcp*{Signal to stop}

\ForEach{$\text{queue} \in \text{MLQ}$}{
    \tcp{Skip lower-priority queues if higher ones exist}
    \If{\text{stop}}{\textbf{break} \;} 
    
    \While{$\text{queue}$ is not empty}{
        $\tau \gets \text{queue.peek()}$ \tcp*{Peek at the front trajectory}
        
        \tcp{Step 1: Find all candidate instances}
        $\text{candidates} \gets \emptyset$ \;
        \ForEach{instance $i \in S$}{
            \If{$\text{check\_routable}(S, \tau, i)$}{
                $\text{candidates} \gets \text{candidates} \cup \{i\}$ \;
            }
        }
        
        \If{$\text{candidates} = \emptyset$}{
            $\text{stop} \gets \text{True}$ \;
            \textbf{break} \;
        }
        
        \tcp{Step 2: Group candidates by instance model version (ascending)}
        $\text{groups} \gets \text{group\_by\_ascending\_version}(\text{candidates}, S)$ \;
        
        \tcp{Step 3: Compute ideal throughput gain}
        $\Delta \mathcal{T}_{\text{ideal}} \gets \frac{1}{k_1 \times k_5 \times \tau.\text{length} + \max(k_2, k_3) + k_4}$ \;
        
        \tcp{Step 4: Waterfall model selection}
        $\text{selected} \gets \text{None}$ \;
        \ForEach{$\text{group} \in \text{groups}$}{
            $\Delta \mathcal{T}_{\text{max}} \gets -\infty$ \;
            $\text{best\_inst} \gets \text{None}$ \;
            
            \ForEach{instance $i \in \text{group}$}{
                $\Delta \mathcal{T}_i \gets \text{compute\_marginal\_gain}(S, \tau, i)$ \;
                \If{$\Delta \mathcal{T}_i > \Delta \mathcal{T}_{\text{max}}$}{
                    $\Delta \mathcal{T}_{\text{max}} \gets \Delta \mathcal{T}_i$ \;
                    $\text{best\_inst} \gets i$ \;
                }
            }
            
            \If{$\Delta \mathcal{T}_{\text{max}} \ge \mu \times \Delta \mathcal{T}_{\text{ideal}}$}{
                $\text{selected} \gets \text{best\_inst}$ \;
                \textbf{break} \tcp*{Accept (above the threshold)}
            }
        }
        
        \If{$\text{selected} \neq \text{None}$}{
            \tcp{Step 5: Route trajectory}
            $\text{routing}.\text{append}((\text{selected}, \{\tau\}))$ \;
            $\text{queue.pop()}$ \tcp*{Remove from queue}
            $S[\text{selected}] \gets \text{update\_snapshot}(S[\text{selected}], \tau)$ \;
        }
        \Else{
            $\text{stop} \gets \text{True}$ \;
            \textbf{break} \tcp*{No instance meets threshold}
        }
    }
}

\Return $\text{routing}$ \;
\end{algorithm}

\clearpage

\begin{algorithm}[!h]
\caption{Synchronization strategy.}
\label{alg:sync}
\small
\KwIn{$S$: snapshot of rollout instances, $\text{ts\_trajs}$: trajectories in TS, 
$\text{ps\_version}$: current PS model version}
\KwOut{List of instances for synchronization}
$\text{sync} \gets \emptyset$ \tcp*{Instances to synchronize}
$\text{candidates} \gets \emptyset$ \tcp*{Candidate instances to synchronize}

\tcp{Step 1: Identify eligible instances}
\ForEach{instance $i \in S$}{
    \If{$\text{ps\_version} > S[i].\text{inst\_version}$}{
        \tcp{Check if any trajectory can be routed to instance i under the current version}
        $\text{can\_route} \gets \text{False}$ \;
        \ForEach{trajectory $\tau \in \text{ts\_trajs}$}{
             
            \If{$\text{check\_routable}(S, \tau, i)$}{
                $\text{can\_route} \gets \text{True}$ \;
                \textbf{break} \;
            }
        }
        \If{not $\text{can\_route}$}{
            $\text{candidates} \gets \text{candidates} \cup \{i\}$ \;
        }
    }
}

\tcp{Step 2: Tentative update for each candidate}
\ForEach{instance $i \in \text{candidates}$}{
    \tcp{Create tentative snapshot with updated version}
    $S_{{temp}} \gets S$ \;
    $S_{{temp}}[i].\text{inst\_version} \gets \text{ps\_version}$ \;
    
    \tcp{Simulate routing on the tentative snapshot}
    $\text{routing} \gets \text{RoutingStrategy}(S_{{temp}}, \text{ts\_trajs})$ \;
    
    \tcp{Check if any trajectory would be routed to this instance}
    $\text{routed\_to\_i} \gets \text{False}$ \;
    \ForEach{$(\text{inst}, \text{trajs}) \in \text{routing}$}{
        \If{$\text{inst} = i$}{
            $\text{routed\_to\_i} \gets \text{True}$ \;
            \textbf{break} \;
        }
    }
    
    \If{$\text{routed\_to\_i}$}{
        $\text{sync}.\text{append}(i)$ \;
    }
}

\Return $\text{sync}$ \;
\end{algorithm}

\newpage

\begin{algorithm}[!h]
\caption{Migration strategy.}
\label{alg:migration}
\small
\KwIn{$S$: snapshot of rollout instances}
\KwOut{List of (instance, trajectories) pairs for migration}

$\text{migration} \gets \emptyset$ \tcp*{Migration list}

\tcp{Case 1: Handle instances with excessive waiting trajectories}
\ForEach{instance $i \in S$}{
    $\text{wait\_cnt} \gets |S[i].\text{wait\_trajs}|$ \;
    \If{$\text{wait\_cnt} > \varphi_{\text{wait}}$}{
        $\text{excess} \gets \text{wait\_cnt} - \varphi_{\text{wait}}$ \;
        $\text{mig\_trajs} \gets \text{excess trajectories from } S[i].\text{wait\_trajs}$ \;
        $\text{migration}.\text{append}((i, \text{mig\_trajs}))$ \;
        $S[i].\text{wait\_trajs} \gets S[i].\text{wait\_trajs} \setminus \text{mig\_trajs}$ \;
    }
}

\tcp{Case 2: Handle excessive throughput imbalance between instances}
$\text{throughputs} \gets \emptyset$ \;
\ForEach{instance $i \in S$}{
    $\text{throughputs}[i] \gets \mathcal{T}_{i}(S)$ \tcp*{Compute throughput using cost model}
}

$\text{max\_inst} \gets \arg\max_{i} \text{throughputs}[i]$ \;
$\text{min\_inst} \gets \arg\min_{i} \text{throughputs}[i]$ \;
$\text{gap} \gets \frac{\text{throughputs}[\text{max\_inst}]}{\text{throughputs}[\text{min\_inst}]}$ \;

\If{$\text{gap} > \varphi_{\text{throughput}}$}{
    $\text{all\_trajs} \gets S[\text{max\_inst}].\text{run\_trajs} \cup S[\text{max\_inst}].\text{wait\_trajs}$ \;
    
    \tcp{Remove trajectories already handled in Case 1}
    \ForEach{$(\text{mig\_inst}, \text{mig\_trajs}) \in \text{migration}$}{
        \If{$\text{mig\_inst} = \text{max\_inst}$}{
            $\text{all\_trajs} \gets \text{all\_trajs} \setminus \text{mig\_trajs}$ \;
        }
    }
    
    \If{$\text{all\_trajs} \neq \emptyset$}{
        $\text{migration}.\text{append}((\text{max\_inst}, \text{all\_trajs}))$ \;
    }
}

\Return $\text{migration}$ \;
\end{algorithm}